\PassOptionsToPackage{table,xcdraw}{xcolor}
\documentclass[letterpaper,twocolumn,10pt]{article}
\usepackage{usenix}
\usepackage{subfig}
\usepackage{color}
\usepackage{babel,blindtext}
\usepackage{tikz}
\usepackage{amsmath}
\usepackage{comment}
\usepackage{mathtools}
\usepackage{booktabs}
\usepackage{lipsum}
\usepackage{listings}
\usepackage{xspace}
\usepackage{listings}
\usepackage{bbding}
\usepackage{pifont}
\usepackage{wasysym}
\usepackage{amssymb}
\usepackage{amsmath}
\usepackage{amsfonts}
\usepackage{amssymb}
\usepackage{caption}
\usepackage{setspace}
\usepackage{float}
\usepackage{algorithm}
\usepackage{algpseudocode}
\usepackage{multirow}
\usepackage{cleveref}
\usepackage{fancyhdr}
\usepackage[utf8]{inputenc}
\usepackage{amsfonts}
\usepackage{amssymb}
\usepackage{graphicx}
\usepackage[symbol]{footmisc}
\usepackage{enumitem}
\usepackage{tikz}
\usepackage{subfig}
\usepackage{twoopt}
\usetikzlibrary{fit}

\usepackage{array}
\usepackage{booktabs}
\usepackage{multirow}

\NewDocumentCommand{\rot}{O{45} O{1em} m}{\makebox[#2][l]{\rotatebox{#1}{#3}}}%

\newcommandtwoopt\Textbox[5][0.9cm][2cm]{%
\begin{tikzpicture}[remember picture,overlay]
  \coordinate (aux) at ([xshift=#1]#4);
  \node[inner ysep=3pt,yshift=0.6ex,draw=darkgreen,
    fit=(#3) (aux),baseline] 
    (box) {};
  \node[text width=#2,anchor=north east,
    font=\sffamily\footnotesize,align=right] 
    at (box.north east) {#5};
\end{tikzpicture}%
}
\DeclareCaptionSubType*{algorithm}

\DeclareCaptionLabelFormat{alglabel}{Alg.~#2}

\definecolor{verde}{rgb}{0.25,0.5,0.35}
\definecolor{myyellow}{RGB}{229, 148, 12}
\definecolor{jpurple}{rgb}{0.5,0,0.35}
\definecolor{darkgreen}{rgb}{0.0, 0.2, 0.13}
\definecolor{forestgreen}{rgb}{.0,.60,.0}
\definecolor{oldmauve}{rgb}{0.4, 0.19, 0.28}
\definecolor{blueannoback}{RGB}{234,242,250}
\newcommand{\mycode}{
\lstset{
    language=Java,
    basicstyle=\ttfamily\small,
    keywordstyle=\color{blue}\bfseries,
    keywordstyle=[2]\color{blue}\bfseries,
    keywordstyle=[3]\color{jpurple}\bfseries,
    deletekeywords = {return, if, public, private},
    morekeywords=[2]{function, if, public, private, internal, external, balanceOf},
    morekeywords=[3]{address, IERC20, uint, string, bytes, msg.sender, to, from},
    escapeinside={\%*}{*)},  
    stringstyle=\color{oldmauve},
    commentstyle=\color{red},
    morecomment=[s][\color{darkgreen}]{/*}{*/},
    morecomment=[s][\color{red}]{@}{;},
    extendedchars=true,
    showspaces=false,
    showstringspaces=false,
    numbers=left,
    numberstyle=\tiny,
    breaklines=true,
    breakautoindent=true,
    captionpos=b,
    xleftmargin=0pt,
    tabsize=2,
    float,
    floatplacement=!t,
    numbersep=5pt
}}

\newcommand{\ie}{i.e., }
\newcommand{\eg}{e.g., }

\newcommand{\efcf}{\textsc{EF/CF} \cite{efcf}}

\newcommand{\optik}{\textsc{Optik} \cite{optik}}
\newcommand{\optiknq}{\textsc{Optik}}

\newcommand{\smartestnq}{\textsc{SmarTest}}

\newcommand{\manticorenq}{\textsc{Manticore}}

\newcommand{\sailfish}{\textsc{Sailfish} \cite{sailfish}}
\newcommand{\sailfishnq}{\textsc{Sailfish}}

\newcommand{\mau}{\textsc{Mau}\cite{Mau}}
\newcommand{\maunq}{\textsc{Mau}}
\newcommand{\maiannq}{\textsc{Maian}}
\newcommand{\teether}{\textsc{TeEther} \cite{teether}}
\newcommand{\teethernq}{\textsc{TeEther}}
\newcommand{\ilf}{\textsc{ILF} \cite{learnedfuzzing}}
\newcommand{\ilfnq}{\textsc{ILF}}

\newcommand{\ethbmcnq}{\textsc{EthBMC}}

\newcommand{\echidna}{\textsc{Echidna} \cite{echidna}}
\newcommand{\echidnanq}{\textsc{Echidna}}

\newcommand{\solseenq}{\textsc{SolSee}}

\newcommand{\sfuzz}{\textsc{sFuzz} \cite{sfuzz}}

\makeatletter
\def\RemoveSpaces#1{\zap@space#1 \@empty}
\makeatother

\def\sys{\textsc{SmartSys}\xspace}
\begin{document}

%don't want date printed
\date{}
%make title bold and 14 pt font (Latex default is non-bold, 16 pt)
%High-Coverage and Scalable Smart Contract Testing via Modularized Multimodal Learning on Hybrid Fuzzing
\title{Detecting Buggy Contracts via Smart Testing}
\author{Sally Junsong Wang$^*$, Jianan Yao$^*$, Kexin Pei$^{\dagger}$, Hidedaki Takahashi$^*$, Junfeng Yang$^*$
\\  $^*$Columbia University, NY \xspace
    $^{\dagger}$The University of Chicago, IL\\%\\
{}}
\maketitle
\begin{abstract}
Smart contracts are susceptible to critical vulnerabilities. Hybrid dynamic analyses, such as concolic execution assisted fuzzing and foundation model assisted fuzzing, have emerged as highly effective testing techniques for smart contract bug detection recently. This hybrid approach has shown initial promise in real-world benchmarks, but it still suffers from low scalability to find deep bugs buried in complex code patterns. We observe that performance bottlenecks of existing dynamic analyses and model hallucination are two main factors limiting the scalability of this hybrid approach in finding deep bugs. 

To overcome the challenges, we design an interactive, self-deciding foundation model based system, called \sys, to support hybrid smart contract dynamic analyses. The key idea is to teach foundation models about performance bottlenecks of different dynamic analysis techniques, making it possible to forecast the right technique and generates effective fuzz targets that can reach deep, hidden bugs. To prune  hallucinated, incorrect fuzz targets, \sys feeds  foundation models with feedback from dynamic analysis during compilation and at runtime.

%Smart contracts are increasingly susceptible to critical vulnerabilities. While fuzzing is widely used to find bugs in smart contracts, it often struggles to generate the specific inputs necessary to trigger deep bugs. Hybrid fuzzing, which combines fuzzing with concolic execution, offers partial improvements in coverage, but concolic execution is computationally expensive and does not fully resolve the coverage problem. It remains unclear when to apply concolic execution and how to generate effective inputs to explore deep, hard-to-reach code paths.

%To address these challenges, we present \sys, a self-deciding smart contract analysis framework that leverages a foundation model as a forecast mechanism. After initial fuzzing, the forecast model analyzes coverage data to predict whether concolic execution or foundation-model-assisted fuzzing will best improve coverage for underexplored functions. Additionally, the foundation model, guided by real-time compiler feedback and fuzzing coverage data, generates deep-bug-revealing fuzz targets by constructing effective function call sequences while suppressing hallucination.

The interesting results of \sys include: i) discovering a smart contract protocol vulnerability that has escaped eleven tools and survived multiple audits for over a year; ii) improving coverage by up to 14.3\% on real-world benchmarks compared to the baselines.
\end{abstract}

\section{Introduction}
\label{sec:introduction}

Smart contracts are computer programs facilitating trustless transactions on blockchain. As of August 31, 2024, over 67 million smart contracts with a locked value of 47 billion US dollars have been deployed on Ethereum~\cite{defillama}. Unfortunately, smart contracts' widespread adoption has also exposed the detrimental impact of bugs, frequently leading to substantial financial losses. In July and August of 2024 alone, 379 million US dollars were lost due to exploited smart contract bugs across 15 documented incidents~\cite{defi_rekt}.

High-profile smart contract bugs have motivated a wide range of dynamic analysis techniques \cite{learnedfuzzing, verisol, teether, sailfish}. 
As in traditional software, fuzzing has been used to find bugs in smart contracts~\cite{echidna, mythril, contractfuzzer, harvey, efcf}. 
However, fuzzing often struggles with complex path constraints, leaving bugs in rarely-reached branches undetected. 
While concolic execution-assisted hybrid fuzzing has shown promise in solving complex constraints, it still faces a considerable challenge in how to optimally allocate resources between fuzzing and concolic execution in the hybrid setting.

Concolic execution is a computationally expensive process that requires intensive usage of CPU and memory to track and solve path constraints. Consequently, employing concolic execution universally in hybrid fuzzing is undesirable. 
However, it is uncertain whether concolic execution can uncover new program paths missed by fuzzing until we invest significant computational resources running both methods.
% \jianan{Do we want to discuss the second challenge of concolic execution, the challenge to accurately model the execution environment in the EVM world? I personally feel it may be distractive in the introduction.}
% \kexin{I think it's fine to just stick with concolic execution is expensive in the intro, so we have to be cautious to do forecasting}

In addition to resource constraints and limited scalability, new studies show that bugs are often deeply embedded in execution paths \cite{arbiter}, and more importantly, detecting such ``deep bugs'' requires writing fuzz targets that can correctly capture the underlying transactional logic.
Developing bug-revealing fuzz targets can require precise function invocation sequences informed by the underlying business logic and blockchain context, demanding developers to invest hours, if not days, in the process.

To tackle these challenges, we introduce \sys, an interactive, self-deciding foundation model system designed to support hybrid fuzzing by addressing the issues of when to use concolic execution v.s. fuzzing and how to generate effective fuzz targets.
\sys' interactive models consist of a forecast model deciding whether to invoke concolic execution or a generator model to produce correct and context-aware fuzz targets. Foundation models (or interchangeably, large language models) are particularly well-suited for these tasks, because they have large search space thanks to pre-training and capabilities in code understanding and generation for specific transactional contexts.

\sys differs from prior foundation model assisted frameworks \cite{llmsymbolic, llm4fuzz, llm4protocolfuzzing, llm4uzzingsurvey} in its unique, two-level interactions: i) the interactions between two  foundation models to improve scalability of dynamic analysis; ii) the interactions between foundation models and dynamic analysis to facilitate correct fuzz targets generation.
A technique is most effective when it incorporates domain-specific knowledge to automatically detect bugs that fall within its strength. The challenge is to know in advance which code segments of a smart contract is best suited for fuzzing or concolic execution. 
%\junfeng{what are the techniques to pick from? randomized fuzzing vs concolic execution, right? If so we should be explicit}.
To that end, we introduce a forecast model fine-tuned on coverage reports from dynamic analysis. Foundation models are suited for forecasting, because they can learn domain-specific information about smart contracts via fine-tuning and then generate predictions.

To write correct and effective fuzz targets, we observe that foundation models are receptive to feedback from compilers and other techniques. Feedback from compilers, such as error messages, and from other techniques, such as coverage data, enhances foundation models' code generating ability and enables them to automatically fix the errors of generated fuzz targets with concrete error points and transactional contexts. A key difference between our fuzz target correction approach and existing code editing models such as Codey~\cite{codey} is the level of feedback and automation: We let the model receives feedback during compile time and runtime simultaneously and automatically without human reviews.

Evaluation on real-world smart contract benchmarks demonstrates that \sys improves coverage by up to 14.3\%. Interestingly, \sys has discovered a critical vulnerability that has escaped eleven existing tools for more than a year. Anonymized reproducible artifacts are available\footnote{https://anonymous.4open.science/r/SCTest-5F4C/algo/executor/corpus}.

\mycode
\begin{lstlisting}[float,floatplacement=H,basicstyle=\fontsize{8}{9}\selectfont, caption={{P1. Common bottleneck for fuzzing: CVE-2023-34234 (simplified for readability).}}, label={fuzzingexample},numbers=left, xleftmargin=2em]
function castVote(uint id, address voter, string calldata reason, bytes memory params. bytes memory sig) external returns (bytes) {
  //challenge: magic number constraints
  if (abi.encode(voter, id)==keccak256(abi.encode(bytes(reason), params, sig))) {
      _castVoteInternal(...); //buggy call
  }
  revert()
}
...
\end{lstlisting}

In summary, this paper makes the following contributions:

\begin{itemize}[leftmargin=*]
\item To the best of our knowledge, \sys is the first interactive, self-deciding system that achieves high scalability while generating hallucination-free fuzz targets to detect ``deep bugs'' in real-world smart contracts.
\item \sys taps into foundation models'  forecasting power via augmented code understanding ability, generating test cases that consistently improve \efcf's coverage, an AFL++ adapted fuzzer for Solidity, on sampled 82 projects.

\item We present a trio of comprehensive benchmarks useful for fine-tuning, coverage, and bug detection analyses. The benchmark is comprised of the top five types of commonly exploited zero-day vulnerabilities from 2022-2024. 
\end{itemize}
\section{Motivation}
\label{sec:motivation}

While much research has devoted efforts to various techniques for smart contract bug detection~\cite{verx, smartest, verismart, echidna, optik}, 
%\jianan{What are "each techniques"?}, 
they do not offer an automated and generic solution for testing real-world smart contracts with high scalability and high coverage. 
In this section, we first discuss the common challenges faced by existing techniques and introduce our solutions.
We then use a recent hack \cite{velocorhack, velocoresource} causing \$6.8 million loss to the Ethereum Protocol Velocore to concretely show how the discussed challenges manifest.

\subsection{Common Bottlenecks}
\label{subsec:coomon_bottlenecks}

%\junfeng{overall, I like the structure of this section to highlight the research ideas in \sys. However, there's not much discussion in later section on the loop invariant inference and the different types of bugs. Why is that?}

We have studied eleven state-of-the-art tools \cite{contractfuzzer, teether, oyente, echidna, optik, ethbmc, verismart, verisol, verx, mythril, slither} and identified three key challenges. 
In the following, we illustrate each challenge using real-world code patterns from high-impact smart contract bugs registered in Common Vulnerability and Exposure (CVE) database and Code4Rena \cite{code4rena}, a smart contract auditing platform.

\begin{sloppypar}

\vspace{0.1cm}\noindent\textbf{P1. Fuzzing Coverage Plateaus.} Fuzzing can quickly explore input space and generate test cases for loosely constrained input, such as \texttt{if(input>10)}. However, fuzzing frequently hits coverage plateaus on complex magic number constraints, such as \texttt{if(input==0xdeadbeef)} and \texttt{if(input==keccak256(0xdeadbeef))}.
%\junfeng{this is not a valid hex number (digits + abcdef)}, \junfeng{invalid hex number. make sure the sample code + some boilerplate compiles}
Due to heavy dependence on cryptography in real-world smart contracts, magic number constraints are highly prevalent. An even more damaging ripple effect is that exploitable bugs are often hidden deep under magic number constraints, curtailing fuzzers' effectiveness. 
\end{sloppypar}

To show this challenge, we ran existing fuzzers~\cite{learnedfuzzing, sfuzz, efcf, echidna} that implement a wide range of coverage-guided optimizations on Listing~\ref{fuzzingexample} with a generous time budget of 5 hours. 
None could escape the coverage plateau at line 3 caused by a magic number constraint. Learning-based fuzzer \cite{learnedfuzzing} and greybox fuzzers \cite{echidna, efcf} hit plateaus whenever hashing-related magic number constraints are present. To generate inputs that can move beyond the constraint at line 3, there exist only limited combinations of \texttt{reason}, \texttt{param}, and \texttt{sig} that can hash to the concrete value on the left side.
%\junfeng{this is not valid assumption. typically hashing functions are shrinking, mapping larger domain to a smaller one.}
Assuming the default size of \texttt{bytes} type is 32, the hash function is $2^{\frac{n}{2}}$ collision resistant, and the left side is a concrete value, the possibility of a fuzzer generating such input by random mutation is $\frac{1}{2^{\frac{2^{32}*2^{32}*2^{32}}{2}}}$, roughly one in billions.
%\junfeng{since there are more than one combinations that hash to the magic number, this probability calculation is incorrect.} 
% \begin{scriptsize}
 \mycode
\begin{lstlisting}[float,floatplacement=H,basicstyle=\fontsize{8}{9}\selectfont, caption={P2. Common bottlenecks for concolic execuction: CVE-2023-34459 (simplified for readability).}, label={concolic_example},numbers=left, xleftmargin=2em]
function validate(address proposer, string memory key) internal returns (uint) {
  uint len = bytes(key).length; 
  //challenge 1: loop
  for (uint i=0; i<len; ++i) {
    uint8 x = uint8(bytes(key)[i]); 
    //challenge 2: non-linear constraints
    uint c = x*x*x-12;
    if (0 < c && c < 16) {
      //bug haappens here
    }   
  }
  return 0; 
}
\end{lstlisting}
% \end{scriptsize}

\vspace{0.1cm}\noindent\textbf{P2. Slow Concolic Execuction.} Listing~\ref{concolic_example} highlights the two challenges faced by existing symbolic or concolic execution engines for smart contracts: loops and non-linear constraints. For each iteration of the loop at line 4, it creates a conceptual fork of the branching point at line 8. Each branching point has two possible states and adds two constraints to the solver. Given that the loop assumes at least 40 iterations, that's $2^{40}$ states and constraints to be solved. Although solutions to these challenges have been proposed in the past \cite{dart, backward_symbolic, Klee, teether, ethbmc, solsee}, they either sacrifice completeness, \eg imposing a bound on the number of iterations, or dismiss loops. 

Existing symbolic/concolic execution tools \cite{teether, ethbmc} dismissed this loop, because the paths did not contain pre-defined critical instructions that must contain \texttt{CALLCODE}, \texttt{DELEGATECALL}, \texttt{SELFDESTRUCT} instructions that fit into pre-defined search heuristics.
To investigate how existing tools handle loops, we instrumented Listing~\ref{concolic_example} with required \texttt{CALLCODE} instructions in the contract. 
After modifying the contract to fit pre-defined heuristics, existing tools stopped executing when they hit loop bound of three iterations, thereby sacrificing complete explorations for program behavior. 
Other concolic execution engines for smart contracts, such as \solseenq\xspace and \maiannq, did not handle loops and failed to report results on contracts like Listing~\ref{concolic_example}. 

\begin{figure}[!t]
\centering
\includegraphics[width=\linewidth]{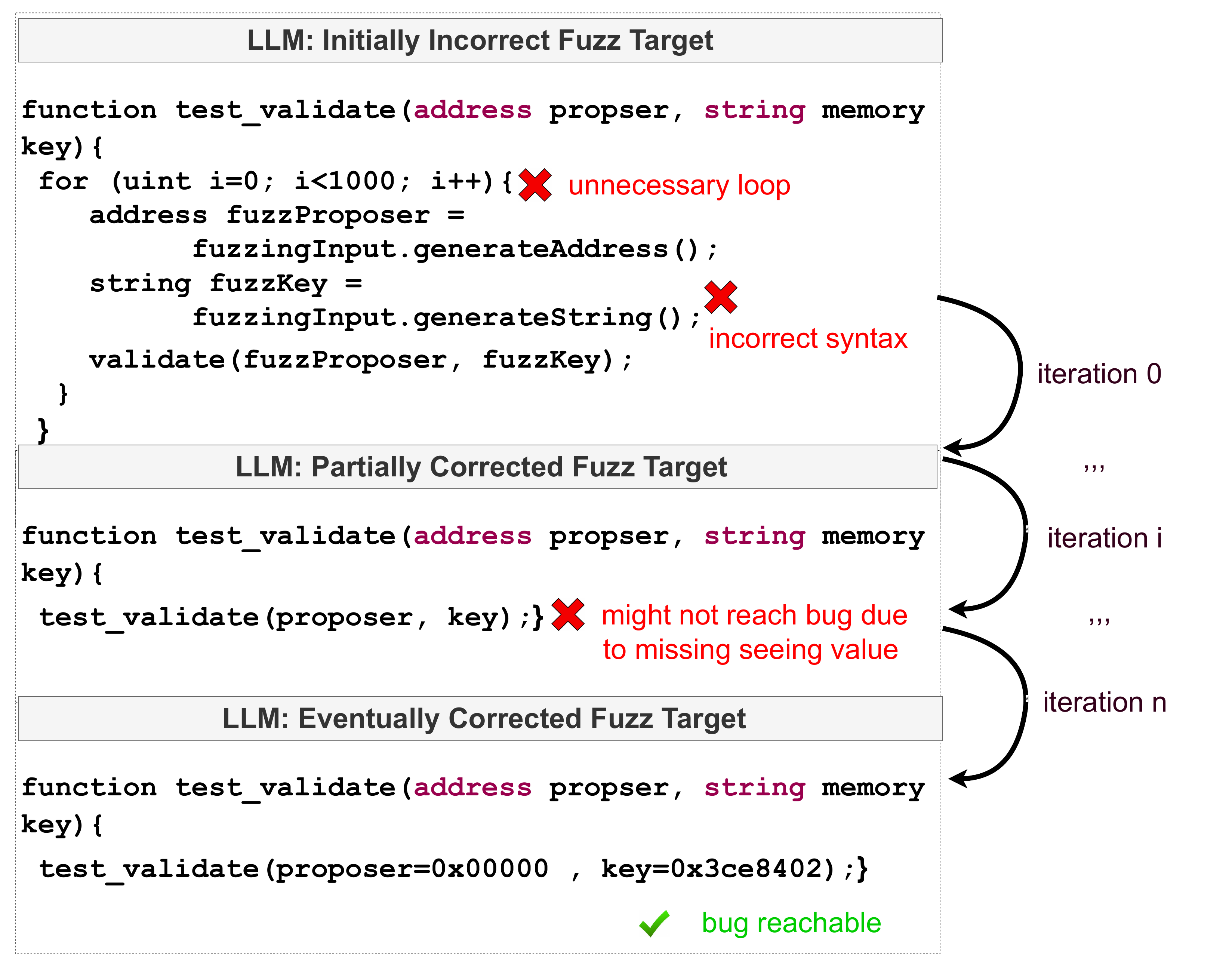}
\caption{Listing~\ref{concolic_example} fuzz targets. Hallucinated fuzz target at the top uses an imaginary loop and incorrect syntax; after a few iterations of fixing, partially corrected fuzz target still cannot reach the bug quickly. Eventually corrected fuzz target is syntactic and semantically correct with seeding values that can overcome plateaus quickly.} 
\label{fig:target2}
\end{figure}

When concolic execution slows down, switching back to fuzzing mode with a good fuzz target can theoretically speed up testing. However, as Fig.~\ref{fig:target2} shows,  LLMs frequently generate incorrect fuzz targets, rendering LLM assisted fuzzing without hallucination suppression as a less attractive alternative. To make things worse, hallucinated fuzz targets in Fig.~\ref{fig:target2} add more futile loops.
%\junfeng{make sure you observe these problems with some existing concolic execution tools}\kexin{again this does not seem like the challenge of concolic execution}

%\junfeng{so it seems that the challenge is to infer loops, but the solution section is more for making symbolic execution framework faster via implementation tricks.}

\begin{sloppypar}
\vspace{0.1cm}\noindent\textbf{P3. Limited Function Invocation Ordering Analysis.} Prior research \cite{swc, sailfish} indicates that exploitable smart contract bugs are often rooted in specific function call sequences with a malicious input parameter setup. Take Listing~\ref{dataflow} as an example, where each address  is associated with a list of id(s). Naively testing the buggy function by calling \texttt{deposit(from=B, id=1, value=1)} will not work, because \texttt{id2asset}, \texttt{poolBal}, and values will be initialized as zero and fail the assertions at lines 15-16.  
\end{sloppypar}

\begin{sloppypar}
Existing tools unfortunately miss the invocation order. Due to coarse data flow approximation, existing static data flow analysis \cite{slither} cannot generate test cases to find bugs embedded in precise function invocation orders. This limitation arises, because static data flow analysis relies on hand-crafted queries over a graph representation of source code. As mentioned in P2, since the buggy data flow in Listing~\ref{dataflow} does not contain pre-defined instructions for search heuristics, existing symbolic tools miss bugs embedded in such invocation ordering. 
\end{sloppypar}

%\junfeng{this isn't talked in later sections. perhaps fold into test harness generation since to find different types of bugs you need different tests / transaction sequences?}

\subsection{Velocore Hack}
\label{subsec:beanstalk_hack}
We now examine a simplified Velocore hack\footnote{https://github.com/velocore/velocore-contracts/blob/4f9bbbd6e0de052725d8642e3efa4904059d44b2/src/pools/constant-product/ConstantProductPool.sol\#L164} causing a \$6.8 million loss in 2024. An amalgam of bottlenecks identified above culminates in this hack. While existing tools have studied other vulnerabilities \cite{confusum, imitationgame, counterattack, sailfish}, automated detection of the Velocore hack with full code coverage remains an open challenge. 

% \begin{scriptsize}
 \mycode
\begin{lstlisting}[float,floatplacement=H,basicstyle=\fontsize{8}{9}\selectfont, caption={Common bottlenecks for precise function invocation ordering analysis: Code4Rena-2024-04 Report \cite{P3analysis, P3} (simplified for readability).}, label={dataflow},numbers=left, xleftmargin=2em]
uint poolBal;
mapping(uint=>uint) id2asset;
mapping(address=>mapping(uint=>uint)) allowed;
//challenge: specific function invocation order to reach buggy location
function mintDyad (uint id, uint amount) {
  id2asset[id] += amount;
  poolBal += amount;
}

function redeemable (uint id, uint value) {
  allowed[msg.sender][id] = value;
}

function deposit (address from, uint id, uint value) {
  assert(id2asset[id] >= value);
  assert(allowed[from][id] >= value); 
  //buggy line below: wrong computation
  poolBal += value;  //should be poolBal-=value
}


\end{lstlisting}
% \end{scriptsize}

\begin{sloppypar}
 The fee calculation logic bug in Listing~\ref{listing:velocore} is rooted in a specific function invocation order from \texttt{notifyWithdraw} to \texttt{velocore\_execute}. The \texttt{notifyWithdraw} function sets up the value of \texttt{feeMultiplier} from a user-given parameter \texttt{m} and the value of \texttt{lastWithdrawTimestamp} from the current block time. An attacker can pass an arbitrarily high \texttt{m} value to inflate \texttt{feeMultiplier}.  After that, \texttt{velocore\_execute} function computes the attacker controlled effective rate \texttt{effectiveFee1e9} by the formula at line 14.  Given the inflated effective rate, token swapping is performed with \texttt{effectiveFee1e9} favorable to the attacker in the for loop at lines 17-25. Although developers write assertion at line 21 to check the desired range of \texttt{effectiveFee1e9}, coverage bottlenecks prevent automated tools from reaching and triggering the assertion.
\end{sloppypar}

\begin{sloppypar}
\vspace{0.1cm}\noindent\textbf{Analysis Hurdles.} Automating the detection of the fee calculation logic bug with full code coverage is non-trivial. A tool needs to generate an optimal fuzz target that can exercise magic number constraint at lines 12 and 19 (\textbf{P1}), so it can reach the incorrect computation and propagated operations. As discussed in \S\ref{subsec:coomon_bottlenecks}, the code patterns - loops and complex constraints (\textbf{P2}) - at lines 17-25 also cause coverage plateaus in existing fuzzers \cite{contractfuzzer, sfuzz, pluto, learnedfuzzing} and incurs state explosion in existing concolic execution tools \cite{solsee, teether}. 
\end{sloppypar}

\begin{figure}[!t]
\centering
\includegraphics[width=\linewidth]{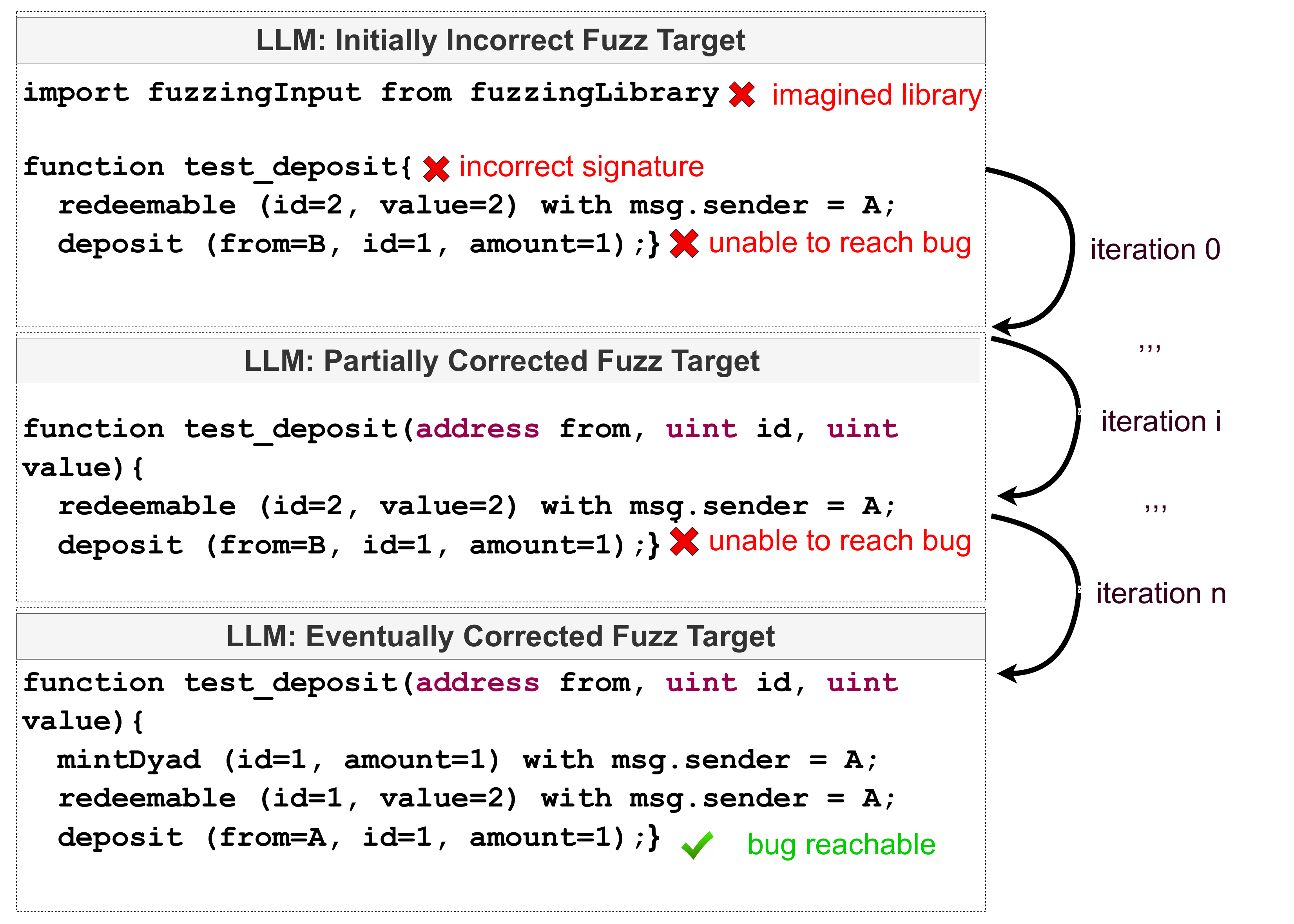}
\caption{Listing~\ref{dataflow} fuzz targets. Hallucinated fuzz target initially includes imaginary libraries and incorrect function signature. After a few iterations of fixing, correct fuzz target at the bottom tells the fuzzer to start with bug reachable seeding values in given function order, and then to mutate input variables in the function deposit via test\_deposit.} 
\label{fig:target3}
\end{figure}

% \begin{scriptsize}
 \mycode
\begin{lstlisting}[float,floatplacement=H,basicstyle=\fontsize{8}{9}\selectfont, caption={Velocore hack causing \$6.8 million loss in June, 2024 (simplified for readability).}, label={listing:velocore},numbers=left, xleftmargin=2em]
contract ConstantProductPool {

  uint lastWithdrawTimestamp, feeMultiplier, fee1e9; 
  
  function notifyWithdraw(uint128 m) {
    feeMultiplier = m; //attacker manipulation
    lastWithdrawTimestamp = uint32(block.timestamp);
  }

  function velocore_execute(Token[] calldata tokens) {
    uint256 effectiveFee1e9 = fee1e9;
    if (lastWithdrawTimestamp == block.timestamp) {
      //propogated wrong calculation
      effectiveFee1e9 = effectiveFee1e9 * feeMultiplier / 1e9;
    }
    //coverage bottleneck
    for (uint256 i = 0; i < tokens.length; ++i) {
      Token token = tokens.uc(i);
      if (token == tokens.uc(123)) {
        swapToken(effectiveFee1e9, ...);
        assert(10/1e9 < effectiveFee1e9 < 1000/1e9);
      } else {
        ...
      }
    }
  }
}
\end{lstlisting}
% \end{scriptsize}
Additionally, to reach the developer-written test oracle at line 21, a tool needs to set up an EVM environment modeling with precise state variable values of \texttt{m} and \texttt{fee1e9} and invokes the two functions - \texttt{notifyWithdraw} and \texttt{velocore\_execute} - sequentially amidst over 100 functions (omitted for simplification) in the contract (\textbf{P3}). This is non-trivial, as state-of-the-art tools \cite{ethbmc, teether, pluto, sailfish, confusum}, including the ones specifically designed for inter-function, inter-contract analysis \cite{ethbmc, sailfish, confusum}, do not consider such invocation ordering, because they are limited by simple inter-function modeling to support expert (manual)-crafted bug search heuristics. As Frank et al.\cite{ethbmc} correctly acknowledges, modeling complex contract behavior can aid the detection of sophisticated bugs in the dark, yet requires substantially higher engineering efforts.

\subsection{Our Solutions}
\label{subsec: solution}

We discuss how \sys addresses the challenges above.

\vspace{0.1cm}\noindent\textbf{Our Solution to P1.} Our solution is driven by the observation that concolic execution and foundation models can solve complex constraints that plateau fuzzing. Yet concolic execution is more expensive than the prompting models in terms of time and memory costs. However, in certain instances, such as mathematical constraint \texttt{if(input==p*r)},
%\kexin{could we refer to your prior hashing example and argue it is still hard to do it, but LLM due to some memorization can still do it?}, 
concolic execution can be more efficient and accurate than foundation models. Given that insight, \sys first uses a fine-tuned forecast model to automatically forecast whether fuzzing-plateaued constraints should be given to concolic execution or to a fuzz target generator model.
%\kexin{still needs to argue why LLM here, is it because the multimodality presented in the code, which LLM is very capable of understanding, happening to be very useful to make such forecast? otherwise there is no need to fine-tune an LLM but just a random small ML model that is much easier to train. better to argue in an organized way in a dedicated subsection 2.4, instead of here addressing each challenge one by one, which does not read super organized.} 
Then \sys re-directs uncovered code according to the forecast model's decision in \S\ref{subsec: fine-tuned Binary Classification Model}.

\vspace{0.1cm}\noindent\textbf{Our Solution to P2.} To reason about loops efficiently, \sys prompts a fuzz target generator model in \S\ref{subsec: Test Case Generation Model} for fuzz targets and if fuzz targets do not overcome the plateau, \sys resorts to concolic execution. As shown in Fig.~\ref{fig:target2}, correct fuzz targets with good seeding parameter values can move beyond the if condition at line 8 in Listing~\ref{concolic_example}: if the parameter setup has \texttt{key[0]=3}, then the first loop iteration will have \texttt{x} = 3 and \texttt{c} = \texttt{x*x*x-12} = 15,
 thus allowing execution beyond \texttt{if(0< c \&\& c <16)} at line 8. 

To handle more complex non-linear constraints, we implemented the first  Directed Automated Random Testing (DART) style concolic execution engine for smart contracts. Although DART \cite{dart} itself is not new, smart contracts require unique blockchain-related optimizations with details in \S\ref{subsec: Optimizing Dynamic Analysis}. 

\begin{sloppypar}
\vspace{0.1cm}\noindent\textbf{Our Solution to P3.} Instead of manually crafting queries on source code data flow, \sys prompts the fuzz target generator model to discover i) potentially malicious input parameter setup and ii) precise, bug-inducive function invocation order as detailed in \S\ref{subsec: Test Case Generation Model}. 

As Fig.~\ref{fig:target3} demonstrates, a correct fuzz target for Listing~\ref{dataflow} should initialize function invocation order with parameter seeding values to trigger the bug: invoking \texttt{mintDyad} and then \texttt{reemable} with \texttt{id=1} from A  will enable \texttt{deposit} fuzzing to pass the assertions at lines 15-16, thereby reaching the bug.
\end{sloppypar} 
To automate hallucination suppression in the fuzz targets generating process in Figs.~\ref{fig:target2} and \ref{fig:target3}, we introduce an iterative algorithm in \S\ref{subsec: hallucination suppression algorithm} that feeds compiler error messages and coverage report back into the fuzz target generator model. The algorithm creates a feedback loop for syntactically correct and bug reachable fuzz targets generation. 

\vspace{0.1cm}\noindent\textbf{Our Solution to Velocore Hack.} \sys employs concolic execution to get past the constraint at line 12 and generates the following fuzz target:
\begin{enumerate} [topsep=3pt,itemsep=-.8ex,partopsep=1ex,parsep=1ex] 
   \item \texttt{fee1e9=1000};
   \item \texttt{notifyWithdraw(m=1000)}; 
   \item \texttt{velocore\_execute(tokens=[123])};  
\end{enumerate}
\begin{sloppypar}
This target works, because when the loop at line 17 unrolls for the first iteration, the value of \texttt{tokens} enables fuzzers to enter the if branch, and \texttt{effectiveFee1e9} = \texttt{effectiveFee1e9} *  \texttt{fee1e9 / 1e9} = 1,000,000 / 1e9 = 0.001, which is larger than the assertion's upper bound at line 21 in  Listing~\ref{listing:velocore} and thus triggers violation. 
\end{sloppypar}
%\junfeng{consider adding a highlevel overview of how our system works on this example and how much better the results are.}
%\input{overview}
\section{Methodology}
\label{sec:methodology}

% Fig.~\ref{Fig:overview} illustrates our efforts to build such an interactive system and the workflow of \sys. Prompts of forecast and generator models are in Appendix~\ref{subsec: finetuning_implementation}.

% \sally{explain step 7}

\subsection{Workflow}
\label{subsec:workflow}
\begin{figure}[!t]
\centering
\includegraphics[width=\linewidth]{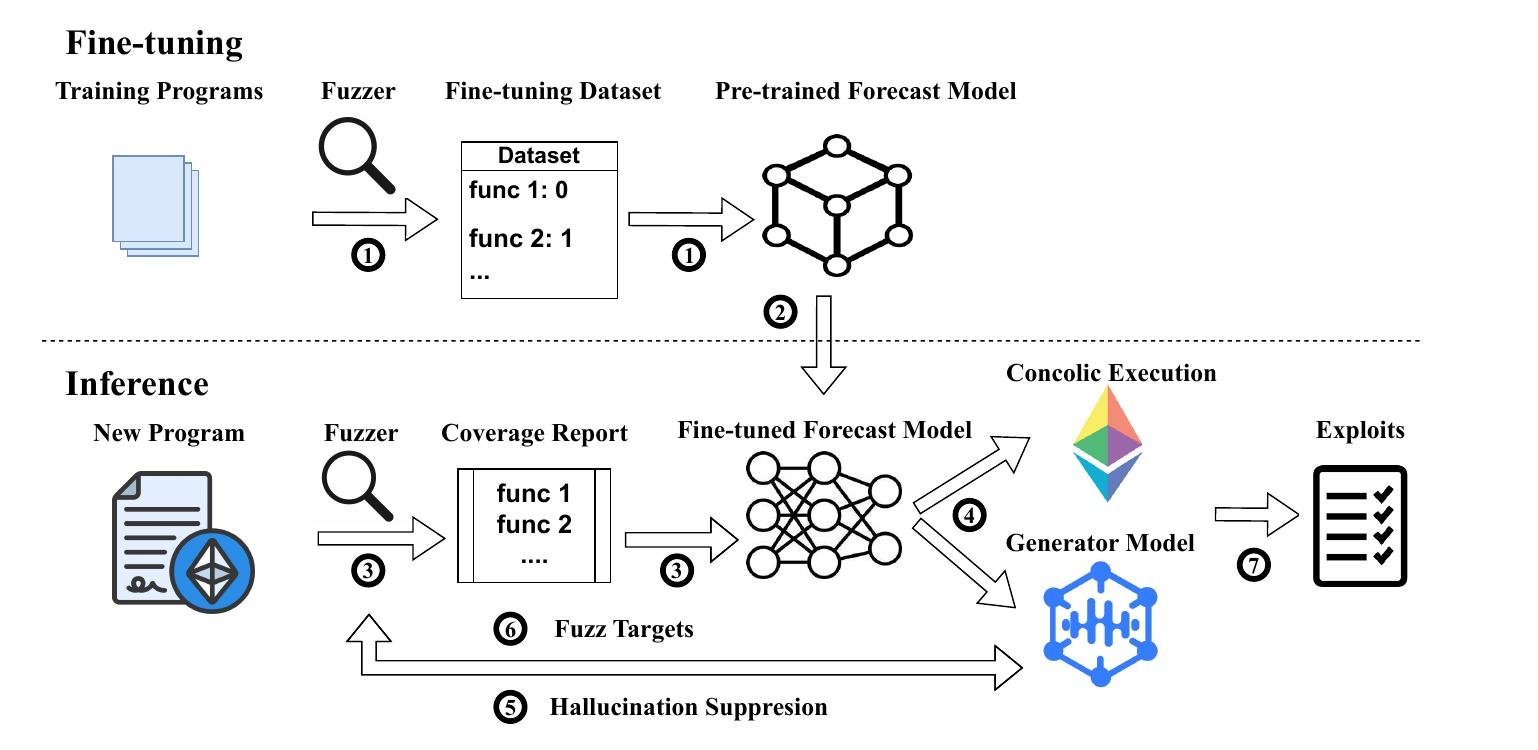}
\caption{\sys's workflow}
\label{Fig:overview}
\end{figure}

Fig.~\ref{Fig:overview} shows the high-level workflow of \sys.
During fine-tuning, the fuzzer generates a fine-tuning dataset from coverage reports (\textcircled{\footnotesize{1}}) using randomized fuzzing and concolic execution.
By comparing the coverage achieved by randomized fuzzing and concolic execution, we create the ground truths to fine-tune the forecast model. 
Once the forecast model is fully fine-tuned (\textcircled{\footnotesize{2}}), it takes as input contract and the coverage report generated by the fuzzer (\textcircled{\footnotesize{3}}), and decides whether to invoke concolic execution or the generator model (\textcircled{\footnotesize{4}}). 
If the generator model is invoked, \sys applies hallucination suppression algorithm (\textcircled{\footnotesize{5}}) and feeds corrected fuzz targets back into the fuzzer (\textcircled{\footnotesize{6}}). \sys repeats this process until coverage plateaus. If concolic execution is invoked, \sys uses concolic execution to overcome plateaus. 
Eventually, \sys generates test cases (possible exploits) as output (\textcircled{\footnotesize{7}}). 
%Specifically, \sys's interactive modules are the following: 

\subsection{Tailored Dynamic Analysis}

\label{subsec: Optimizing Dynamic Analysis}
%\junfeng{why call this fuzzer? It's just fuzzing right?}

\vspace{0.1cm}\noindent\textbf{Fuzzer.} \sys's fuzzer is the starting step of interactions between foundation models and static/dynamic analyses. 
The fuzzer takes as input compiled bytecode
%\kexin{what is compiled smart contract source code? does compiled imply bytecode?} 
and an initial fuzz target constructed from the contract's function signatures and parameters. 
It executes the input contract and generates three outputs: i) a coverage report, where each covered code statement is annotated by *; ii) an initial test case corpus grown from the fuzz target; and iii) a bug report. 
% From compiled smart contracts, \sys first extracts function signatures and parameters to build a fuzz target template in JSON format. 
\sys builds the inter-procedural Control Flow Graph (CFG) that incorporates both the call relations, \eg and edge between function A and B exists if and only if A calls B, and transfers the control induced by the branching statements. 
The coverage report measured on CFG is then used to guide the fuzzer to prioritize test inputs that cover new paths in CFG.

% After extracting function signatures and parameters, \sys builds a control flow graph based on the relations between functions. For example, if function A invokes function B in function A's body, then function B is encoded as a child of function A. Such control flow graphs provide a coarse approximation of program behavior. The benefit of control flow guided fuzzing is that compared to black box (randomized) fuzzing, \sys mutates seeds with control flow guidance. As a result, it can achieve better coverage without significant performance overhead. 

\vspace{0.1cm}\noindent\textbf{Concolic Execution.} 
We implemented a DART (Directed Automated Random Testing) \cite{dart} style concolic execution engine for smart contract analysis with tailored optimizations in \S\ref{subsec: concolic_execution_optimizations}.
Once the forecast model predicts concolic execution should be invoked, our fuzzer seed corpus gets augmented by the input computed by our conclic execution engine.
% The optimization is to find a balance between high coverage (generating test cases that can reach deep code states) and scalability (minimizing state explosion). 

\subsection{Forecast Model}
\label{subsec: fine-tuned Binary Classification Model}

\begin{table}[!t]
\footnotesize
\setlength{\tabcolsep}{3pt}
\centering
\renewcommand{\arraystretch}{1}
\caption{Selected features from fuzzing output and concolic execution traces for forecast model fine-tuning dataset.}
\label{table:binary_training_features}
\begin{tabular}{lll}
\toprule[.6pt]
\textbf{Features} & \textbf{Sources} &\textbf{Rationale} \\ 
\midrule[.6pt]
smart  & source code
   & provide programming contexts \\
 contract & natural language
    & on fuzz targets\\
 \midrule[.6pt]     
coverage  & natural language
    & line by line summary of \\ report & (numbers)
    & code coverage after initial fuzzing \\
\midrule[.6pt]
bug & natural language
    & measure whether test cases\\ report&
    & generated by a module\\
  & & can discover new bugs\\
 \midrule[.6pt] 
un-covered   & source code
    & provide test harnesses for\\
functions & 
    & concolic execution or the\\
    & & generator model\\
\midrule[.6pt]
ground & natural language
    & labeled binary numbers that predict \\ truth & (numbers)
    &  the next test case generating module to \\
  & & escape coverage plateaus\\

\bottomrule[.6pt]
\end{tabular}
\end{table}
The forecast model \textit{self-decides} the optimal strategy for next phase of testing. 
A key challenge of fine-tuning the forecast model is to obtain a sufficiently large and diverse training dataset with useful features in the prompt that present patterns that can generalize to previously unseen contracts. 
We tackle the challenge with the following dataset design and elaborate on the fine-tuning procedure in \S\ref{subsec: compose_interactive_models}.

\vspace{0.1cm}\noindent\textbf{Fine-tuning Dataset.} Table~\ref{table:binary_training_features} summarizes the five features we selected for fine-tuning. 
Specifically, each training sample consists of contract source code, a coverage report that summarizes program statements hitting coverage plateaus, an initial bug report, functions not fully covered after initial fuzzing, and labeled ground truths in 0 (concolic execution) or 1 (generator model). 
The five features are concatenated as a single string, and the model is fine-tuned to predict the next tokens until finishing the last token prediction, \ie the forecast decision.
The top three features of Table~\ref{table:binary_training_features} - smart contract, coverage report, and bug report - are automatically processed as a result of initial fuzzing. 
For the un-covered functions, we have built \texttt{extract\_coverage}, where \sys scans the coverage report and extracts the un-covered or partially covered functions automatically. 

\vspace{0.1cm}\noindent\textbf{Ground Truths.} We automate the ground truth collection by running both the concolic execution and the generator for fuzz target generation. 
After extracting uncovered (or partially covered) functions from fuzzing, we run the concolic execution module and generator model on the contract under test. 
A sample is labeled as 0 if the concolic execution module improves more code coverage and/or finds more bugs than the generator model. 
Otherwise, that sample is labeled as 1.

%\vspace{0.1cm}\noindent\textbf{Labeling Ground Truth.}  In very rare cases, the coverage and bug detection improvement from the concolic execution engine and the generator model are equal. In that instance, the training sample is labeled based on runtime performance. Our experiments show that the forecast model assigns $\sim$86\% un-covered functions to the generator model and the remaining ones to concolic execution. 
 
\subsection{Generator Model}
\label{subsec: Test Case Generation Model}

% \begin{scriptsize}
 \mycode
\begin{lstlisting}[float,floatplacement=H,basicstyle=\fontsize{8}{9}\selectfont, caption={Generating parameter input values as fuzz target.}, label={testharness_1},numbers=left, xleftmargin=2em]
function checkBalance(unit[] tickets, uint amount) external {  
  for (i=0; i<tickets.length; i++){
    if (tickets[i]==amount*amount*amount){
       //do something;   
    } 
  }
}
\end{lstlisting}
% \end{scriptsize}

Once the forecast model decides to take the generator route, \sys prompts the generator model to escape coverage plateaus with generated fuzz targets. 
We observe that the generator is particularly good at generating two types of fuzz targets: i) input parameter values that drive new paths but are blocked by the predicates difficult to solve by the existing solver, and ii) a specific function invocation order that can find new bugs. 

\begin{sloppypar}
\vspace{0.1cm}\noindent\textbf{Input Parameter Values.} 
\sys prompts the generator model to generate input parameter values that can overcome coverage plateaus. 
Take Listing~\ref{testharness_1} as an example: a code snippet that the forecast model predicts to use the generator, because line 2 loop causes concolic execution to significantly slow down. 
\sys first statically extracts the bottleneck constraint at line 3, the partially covered function \texttt{checkBalance}, and input parameters \texttt{tickets} and \texttt{amount} from the coverage report. 
\sys prompts the model in the backend with the extracted function, constraints, and input parameters as input. 
The model generates an output with initialized input parameters: \texttt{checkBalance([8, 1, 1], 2)}, where \texttt{tickets[0]}=8 and \texttt{amount}=2. 
Therefore, during the first iteration of the loop, these seeding values can overcome the constraint at line 3, because \texttt{tickets[0]} = 2*2*2 = 8.
In contrast, obtaining the execution trace for concolic execution becomes expensive due to the loop in line 2, whose number of iterations depends on the length of the external inputs, \ie \texttt{tickets}.
%\junfeng{what do these integers indicate?}

\vspace{0.1cm}\noindent\textbf{Function Invocation Order.} 
\sys also prompts the model to generate specific function invocation orders that can discover hidden logic bugs. 
The prompts used for function invocation order are in Appendix~\ref{subsec: genertor_prompts}.
Function invocation order analysis is driven by smart contracts' unique characteristics: each individual function may be bug-free, but a specific combination of transactions (function invocation orders) can reveal hidden bugs. 
Hybrid fuzzing and concolic execution may achieve 100\% code/path coverage yet fail to detect the logic bugs, because detecting them requires an additional fuzz target - a highly specific function invocation order that can reach the bug. 
\end{sloppypar}

\subsection{Hallucination Suppression Algorithm}
\label{subsec: hallucination suppression algorithm}

\begin{algorithm}[!t]
\footnotesize
\setlength{\tabcolsep}{2pt}
	\caption{Hallucination Suppression Algorithm}
\label{algo: suppress}

\begin{tabular}{lp{2.9in}}
\textbf{Input}:
    & foundation model $M$, fuzzer $F$, smart contract $S$, generated fuzz target $T$,  number of iterative fixes $Itr$ \\

\textbf{Output}:
    &  corrected fuzz target $T'$ \\
\end{tabular} 

\begin{spacing}{1.1}
\begin{algorithmic}[1]

\For{($i$=0; $i$++; $i$ $\leqslant$$Itr$)}
\State Err =  Compile(F(S+T)); \color{blue}\Comment{compiler errors from wrong fuzz targets}\color{black}
\If {Err != None}
\State T = few\_shot\_prompt(M, S+T, Err);
\State T' = T;
\Else
\State Cov =  F(S+T); \color{blue}\Comment{coverage report from fuzzer}\color{black}
\If {Cov improves}
\State break;
\EndIf
\State T' = T;
\State break;
\EndIf
\EndFor
\State return T';
\end{algorithmic}
\end{spacing}
\end{algorithm}

Algorithm~\ref{algo: suppress} summarizes a simple yet highly effective procedure to suppress hallucination by iteratively prompting the foundation model $M$ to fix the error message $err$ from compilers. Once all errors are fixed, $M$ also checks for coverage data improvement $Cov$ from fuzzer. It takes a foundation model $M$, fuzzer $F$, a smart contract $S$, an initially generated fuzz target $T$, and user specified number of iterative fixes $Itr$ as input. If all errors are fixed within $Itr$, the algorithm outputs a correct fuzz target $T'$ as output. Otherwise, $T'$ would be best effort by $M$ within $Itr$. 

The algorithm leverages compiler and coverage report of the fuzzer for the initial fuzz target at lines 2-3, and then the compiler emits error messages with specific line numbers and cause of errors.  The fuzzer reports under-covered lines of code. Then the foundation model is prompted by few-shot examples at line 4 to fix a given fuzz target. The few-shot examples are manually constructed, correct fuzz targets of simple contract programs. 
%\junfeng{what kind of few shot examples do we give here?}
If error persists, the process repeats until the compiler errors are completely fixed or until the algorithm hits user specified iteration limits. 

The types of hallucination fixed by Algorithm~\ref{algo: suppress} include syntactical and semantic errors, as well as bad seeding values or incorrect function invocation order that do not improve coverage. We observe that Algorithm~\ref{algo: suppress} is particularly successful at fixing syntactical and semantic errors, achieving 98\% accuracy on tested fuzz targets, and reasonably successful at improving coverage, generating 56\% correct and coverage-improving fuzz targets.

\section{Implementation}
\label{sec:implementation}

\sys is implemented with a customized concolic execution engine and built on top of \echidna, totaling approximately 9,000 lines of code. 
The forecast model and the fuzz target generator are based on GPT-4o.
% \sys's key innovation is to automatically forecasting the best befitting technique as a next step and composing the interactive forecast and generator models, with GPT-4o as the default. %\jianan{"Concolic execution" comes out of nowhere. Maybe say what "test case generation techniques" are chosen from earlier.}

%To the best of our knowledge, \sys's concolic execution module is the first coverage-maximizing test case generation tool in the smart contract domain. 

\subsection{Trio of Datasets} 
\label{subsec:three_datasets}
%A key aspect of using foundation-model-based approach is to obtain sufficient diverse data for fine-tuning and inference. We collected three datasets for fine-tuning and evaluation. 

Dataset No.1 (\textbf{D1}) contains 60 deployed contracts with complex code patterns reflective of each performance bottleneck in \S\ref{sec:motivation}: i) \emph{benchmark\_1} contains 20 contracts of magic number constraints; ii) \emph{benchmark\_2} contains 20 contracts of loops and nonlinear constraints; iii) \emph{benchmark\_3} contains 20 contracts that require specific function invocation order to reach a buggy program location. This dataset allows us to peak into the effectiveness of proposed techniques on common performance bottlenecks. D1 Dataset is used to answer RQ1 (\S\ref{subsec:RQ1}) and RQ2 (\S\ref{subsec:RQ2}).

Dataset No.2 (\textbf{D2}) contains 60 labeled vulnerable contracts with labeled vulnerabilities from Immunefi \cite{Immunefi} and Code4Rena \cite{code4rena}. In contrast to the benchmarks presented by recent publications and tools \cite{smartbugs, smartest, smartinv, echidna}, D2 Dataset represents the top zero-day vulnerabilities exploited by hackers in newly developed smart contracts. Therefore, D2 Dataset reflects new and pressing vulnerability trends in smart contracts. D2 Dataset is used to answer RQ3 (\S\ref{subsec:RQ3}).

Dataset No.3 (\textbf{D3}) is  used to fine-tune \sys' forecast model. Each bottleneck function and ground truths are automatically labeled.
%\jianan{For file-tuning the forecast model, does each function in each contract in the D3 Dataset have a ground-truth technique choice? If yes, do you manually annotate them?}
D3 Dataset is comprised of 121 real-world contracts sampled from SmartBug Curated Dataset \cite{smartbugs}, the benchmarks of Echidna \cite{echidna} and EFCF \cite{efcf}. In addition to fine-tuning, D3 Dataset is also used to answer RQ4 (\S\ref{subsec:RQ4}).

\subsection{Composition of Interactive Models} 
\label{subsec: compose_interactive_models}
The decision on when to invoke the forecast model and when to invoke the generator model is non-trivial. We observe that naively running fuzzing, concolic execution, and foundation models sequentially degrades the performance more than improves it. 
%\jianan{It is unclear what a strealined pipeline looks like. Does the observation come from some experiments?} 
To solve that challenge, we first implemented an instruction-level coverage monitor for the fuzz scanner and concolic execution. We then implemented a static checker that maps under-covered instructions to under-covered functions and extracts those under-covered functions. 

The under-covered functions from the static checker unlock the key step to composition. The forecast model is fine-tuned with peformance bottleneck functions and forecasting ground truths to decide what to invoke next.
%\jianan{I do not understand "domain-specific" here. Are there different forecast models for different domains?} 
We used 5-fold cross validation on 121 contracts (D3) to construct the training dataset and the holdout test dataset. 
%\jianan{Why do we use Dataset 3, rather than Dataset 1 or 2 or their union for fine-tuning here? We may need to clarify since we claim the benchmarks as one contribution.}
For each fold on average, we split the contracts into 95 training contracts and 26 holdout test contracts.
%\jianan{This is \S\ref{subsec:three_datasets}}).
From the training contracts, we extracted 201 training samples from training contracts that did not achieve full coverage from the prior fuzzing stage. 
%For interested readers, the detailed fine-tuning procedure and prompts are included in Appendix~\ref{subsec: finetuning_implementation}. 

Once the forecast model makes a decision, one may raise another design choice question: If the forecast model initially decides on  concolic execution, why not invoke the forecast or generator model again when the concolic execution engine hits a plateau? Given that concolic execution engine analyzes smart contract programs as a whole, generating new fuzz targets from the generator models adds nothing to concolic execution engine's performance. Another reason is that since the under-covered functions are detailed enough and the forecast model is accurate (92\%), repeatedly invoking forecast or generator models adds a 65-second performance overhead per invocation with minimal coverage and bug detection gains. Therefore, to retain scalability, \sys opts for not invoking the forecast or generator model again when concolic execution engine hits a plateau. 

\subsection{Concolic Execution Optimizations} 
\label{subsec: concolic_execution_optimizations}
\vspace{0.1cm}\noindent\textbf{Ethereum Environment Modeling.} What differentiates our modeling from prior similar works \cite{ethbmc, teether} is that we model precise, transaction-specific EVM environment, including gas usage and cost \cite{gas}, fallback hooks, inter-contract communications, and inter-transaction communications. The benefits of precise modeling are to reduce \sys's false positives and false negatives. 

We model the environment as an abstract EVM World that holds transaction objects and the execution context of a contract. The EVM World takes a list of externally owned accounts, a contract runner for a list of deployed contract addresses, a runtime stack that includes currently executed function calls with associated contract addresses, a transaction queue that includes a list of transactions to be executed, and a fallback monitor to execute fallback functions when required. When execution is completed, the EVM World returns a set of halting instructions, such as \texttt{STOP} and \texttt{RETURN}. If a transaction invokes external contracts, the EVM World returns external call instructions, such as \texttt{CREATE} and \texttt{CREATE2}. 

\vspace{0.1cm}\noindent\textbf{Transaction Level Modeling.} Unlike traditional software, executions of smart contracts' code are not only dependent on input, but also the information on the blockchain. To capture the blockchain environment, we define an account as a tuple \emph{a} = (\textsf{address}, \textsf{balance}). 
we define a transaction as a tuple \emph{t} = (\textsf{function\_call}, \textsf{call\_data}, \textsf{delay}, \textsf{gas}, \textsf{gas\_price}, \textsf{source}, \textsf{destination}, \textsf{value}). Given a contract under test, \textsf{function\_call} is the name of the function to be executed. \textsf{call\_data} are symbolic input variables of \textsf{function\_call}. \textsf{delay} is the network delay of each transaction. \textsf{gas} and \textsf{gas\_price} are the amount of gas and gas price for that transaction. \textsf{source} and \textsf{destination} are the addresses of transaction sender and recipient. \textsf{value} is the WEI value attached to the transaction. A contract has a list of transactions with respect to different function calls and each transaction can be concolically executed parallelly.

\vspace{0.1cm}\noindent\textbf{Concretizing Input-Dependent Loops.} Given a byte array \texttt{B} and an input (\texttt{x})-dependent constraint in the loop, such as \texttt{if (B[i]==x)}, we concretize the byte size of \texttt{B} and the input parameter value. When an array is observed during concolic execution, the engine assumes a value of the input parameter, concretely executes the array byte by byte, and collects constraints from observed executions of concrete values. Given concretized \texttt{B}, the engine assigns the calculated value to the variable in the array \texttt{B} that satisfies constraints involving both \texttt{B} and the input parameter. Details on query optimizations over non-linear constraints are in Appendix~\ref{subsec: extended_concolic_execution}.

\vspace{0.1cm}\noindent\textbf{Coverage Guided Test Case Generation.} Our search heuristic is to schedule states that will likely cover new code. Generated test cases are re-run on compiled smart contract bytecode natively by a test case replay mechanism that we have built from scratch.  Each individual test case points to a concolic execution state. The replay mechanism uses the test cases to create actual transaction objects containing the concrete values used in the test cases. \sys then executes the smart contracts using the concrete transaction objects from test cases to update coverage.

\section{Evaluation}
\label{sec:evaluation}

We evaluate \sys to answer the following questions.
% We include additional ablation study on the choice of models and the decision on fine-tuning in Appendix \S\ref{subsec:ablation}:

\begin{itemize}[leftmargin=*]
\item\textbf{Code Coverage}: How does \sys compare to similar tools on Dataset No.1 (D1) benchmarks in ~\ref{subsec:three_datasets}?

\item\textbf{Effect of Individual Components}:  What's the coverage effect of the hallucination suppression algorithm and interactive models? Among interactive models, what kind of interactive effect do the forecast and generator models produce individually? 

\item\textbf{Bug Detection}: Can \sys effectively detect bugs in recently developed real-world smart contracts in Dataset No.2 (D2) described in ~\ref{subsec:three_datasets}?

\item\textbf{Seeds (Test Case) Quality}: Can \sys generate better initial seeds for other fuzzers? 
\end{itemize}

\vspace{.1cm}\noindent\textbf{Experiments Setup.} 
We selected a wide range of techniques developed in industry and research, including blackbox fuzzing \echidnanq\xspace and concolic execution guided fuzzing \optiknq\cite{optik}, symbolic execution \manticorenq\cite{manticore}, and a learning-based tool \ilfnq\cite{learnedfuzzing}. To ensure as fair comparisons as possible, we did not benchmark \sys against tools in \cite{llm4fuzz, smartest, teether, oyente}, because they either do not optimize for coverage \cite{teether, smartest, oyente} or are not open sourced \cite{llm4fuzz}. We installed and followed the instructions of the latest versions (as of August 24, 2024) of each tool's git repository. 

We ran each tool continuously for 100 iterations and measured coverage at instruction-level granularity. Notably, \manticorenq\xspace does not report instruction-level coverage, so we instrumented its \texttt{manticore.py} file to print out covered instructions at each iteration. Since the other tools reported coverage data, we used their reports. We kept \sys' specifications constant across all runs, search time budget, and memory capacity the same for each benchmark contract, \eg \texttt{--time-budget 180} for fuzzer, \texttt{--max-iterations 10 --depth 2} and 10GB memory budget.

To evaluate individual components, we first ran the baseline concolic execution guided fuzzing with and without the hallucination suppression algorithm. Then we ran the baseline with and without assistance from the forecast and generator models. We measure the delta in the performance gap in code coverage. The fine-grained setup allows for close inspections on the quality of fuzz targets.
It also provides insights into how often the forecast model predicts concolic execution versus the generator model as the next step.

\subsection{RQ1: Code Coverage} 
\label{subsec:RQ1}

\begin{figure*}[!t]
    \centering
    \subfloat[magic number constraints benchmark result]{%
        \includegraphics[width=0.3\textwidth]{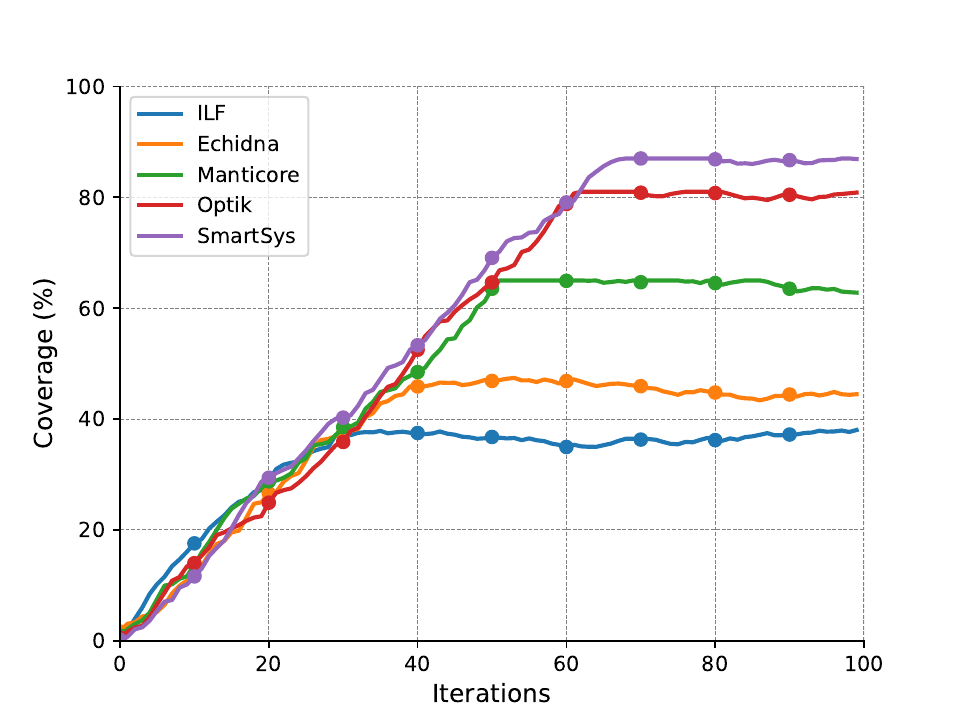} % Replace with your image file
        \label{fig:cov_first_bench}
    }
    \hfill
    \subfloat[loop and non-linear constraints benchmark result]{%
        \includegraphics[width=0.3\textwidth]{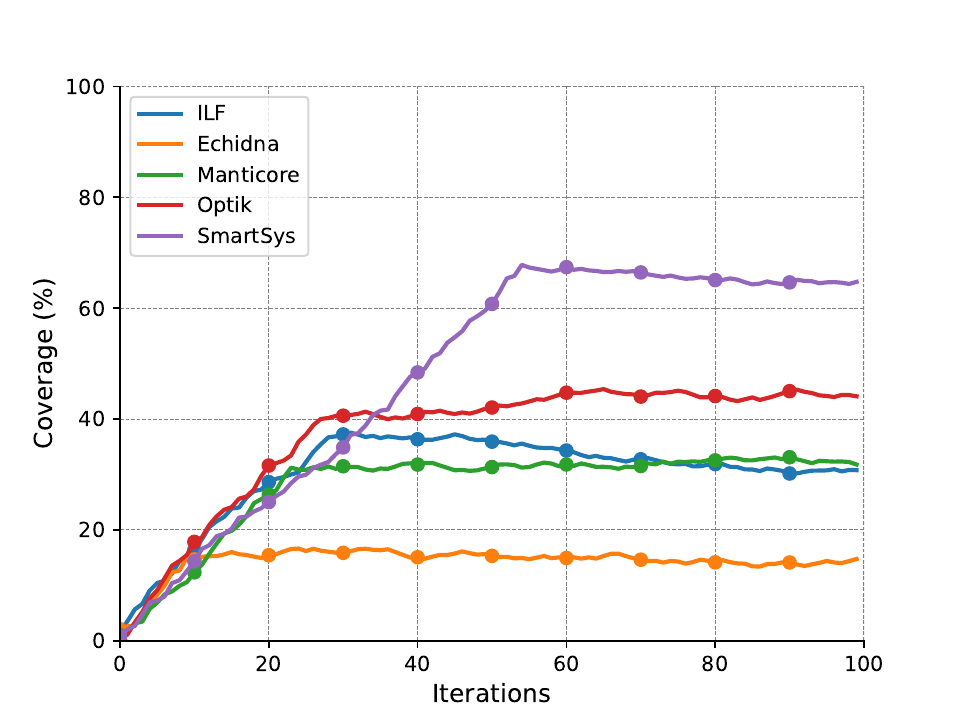} % Replace with your image file
        \label{fig:cov_second_bench}
    }
    \hfill
    \subfloat[specific function invocation order benchmark result]{%
        \includegraphics[width=0.3\textwidth]{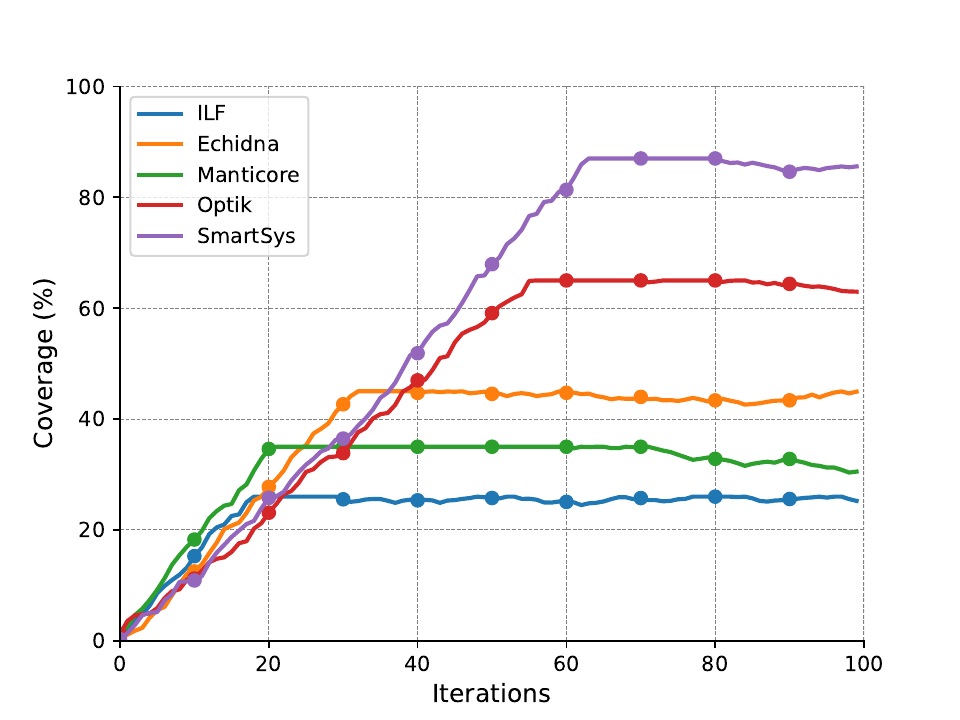} % Replace with your image file
        \label{fig:cov_third_bench}
    }
    \caption{Code coverage comparisons of \sys with similar tools on the three benchmarks of D1.}
    \label{fig:full_coverage}
\end{figure*}

Fig.~\ref{fig:full_coverage} summarizes the average coverage results after three runs across 100 iterations on Dataset No.1 (D1). Across all three benchmarks, \sys outperformed existing tools and reached coverage plateaus later than existing tools by at least 6\%, 13\%, and 12\%. We observed two notable lessons from this experiment. First, concolic execution guided fuzzing, \eg \optiknq, performed better than blackbox fuzzing, \eg \echidnanq, and standalone symbolic execution, \eg \manticorenq, by 26\% and 15\% respectively. Second, interactive models and hallucination-free algorithms further improved the coverage of concolic execution guided fuzzing. 

On the three benchmarks (Figs.~\ref{fig:cov_first_bench}-\ref{fig:cov_third_bench}), \sys obtained coverage at 87\%, 72\% , 86\% and plateaued at about 59, 55, and 61 iterations respectively. By comparison, \optiknq, \manticorenq, \echidnanq, and \ilfnq\xspace obtained coverage at 81\%, 65\%, 55\%, and 56\% at best among all three benchmarks. Hybrid tools, such as \optiknq\xspace and \sys, plateaued later than non-hybrid tools, such as \echidnanq\xspace and \manticorenq, because hybrid tools employ Z3 solvers and coverage-based optimizations to solve complex constraints.

%On the second benchmark (Fig.~\ref{fig:cov_second_bench}), \sys obtained coverage at 72\% and started to plateau at 55 iterations. By comparison, \optiknq, \manticorenq, \echidnanq, and \ilfnq\xspace obtained coverage at 56\%, 38\%, 19\%, and 39\%. Existing tools plateaued before 40 iterations, because randomized inputs from blackbox fuzzers and the runtime cost of solving non-linear constraints blocked further program explorations. \sys' performance was driven by the forecast and generator models, because forecasting relayed code segments to suitable techniques. 

%On the third benchmark (Fig.~\ref{fig:cov_third_bench}), \sys obtained coverage at 87\% and started to plateau at 61 iterations. By comparison, \optiknq, \manticorenq, \echidnanq, and \ilfnq\xspace obtained coverage at 62\%, 43\%, 38\%, and 25\%. \sys had the strongest performance gain on this benchmark, because suppressed hallucination enabled the generator model to generate high-quality fuzz targets. Additionally, while existing techniques relied upon pre-determined heuristics, the generator model's code understanding ability allowed for function invocation ordering analysis in transactional contexts.

\begin{figure*}[!t]
    \centering
    % \hfill
    \subfloat[magic number constraints benchmark result]{%
        \includegraphics[width=0.3\textwidth]{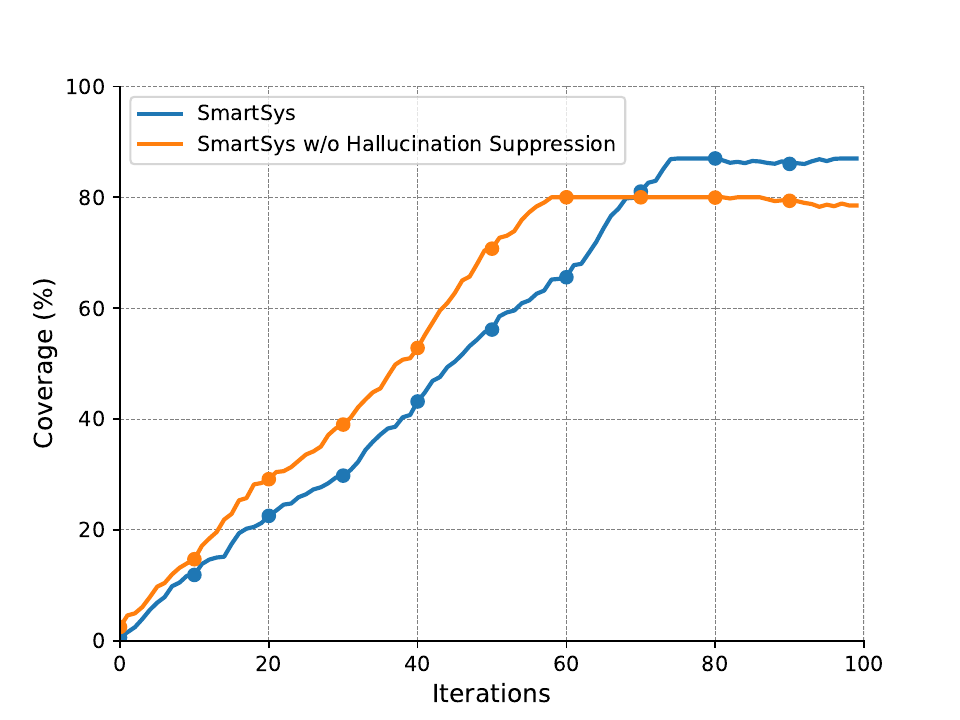} % Replace with your image file
        \label{fig:hallucination_first_bench}
    }
    \hfill
    \subfloat[loop and non-linear constraints benchmark result]{%
        \includegraphics[width=0.3\textwidth]{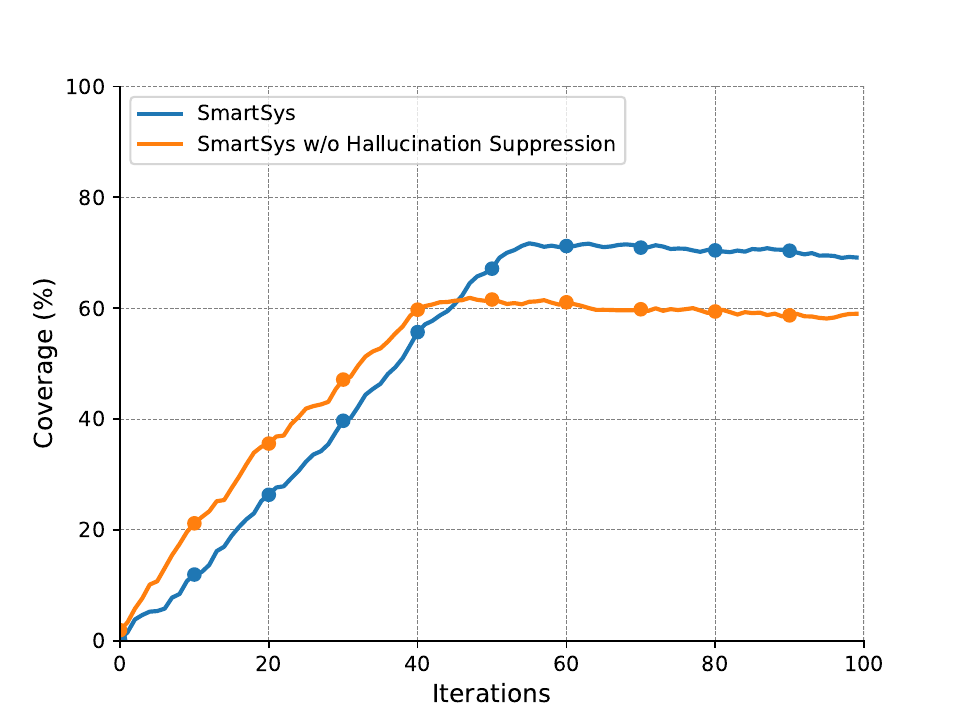} % Replace with your image file
        \label{fig:hallucination_second_bench}
    }
    \hfill
    \subfloat[specific function invocation order benchmark result]{%
        \includegraphics[width=0.3\textwidth]{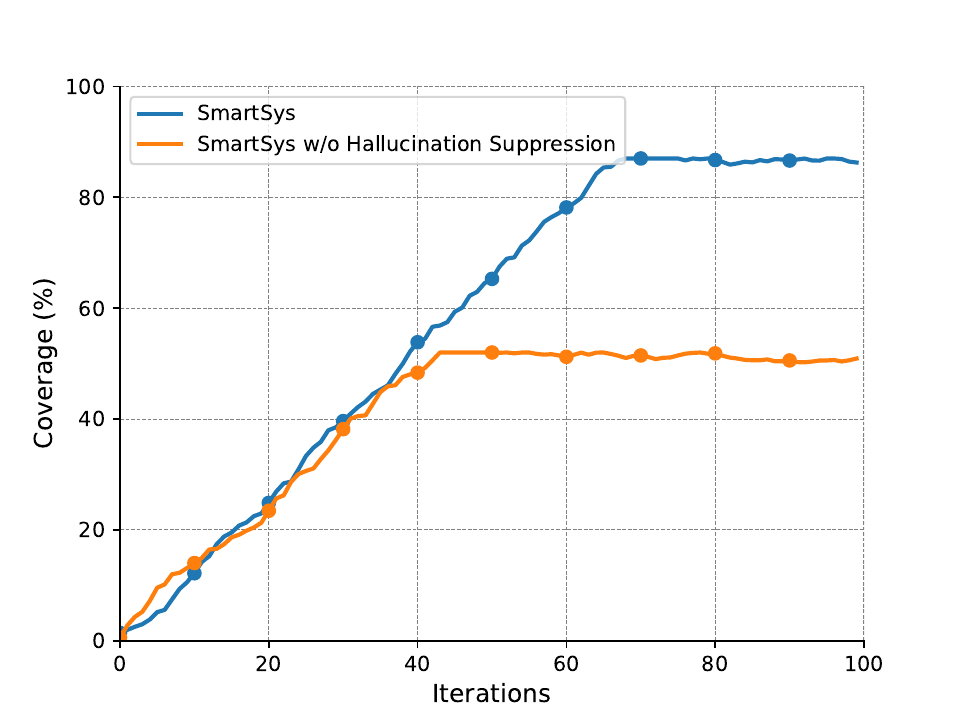} % Replace with your image file
        \label{fig:hallucination_third_bench}
    }
    
    \subfloat[magic number constraints benchmark result]{%
        \includegraphics[width=0.3\textwidth]{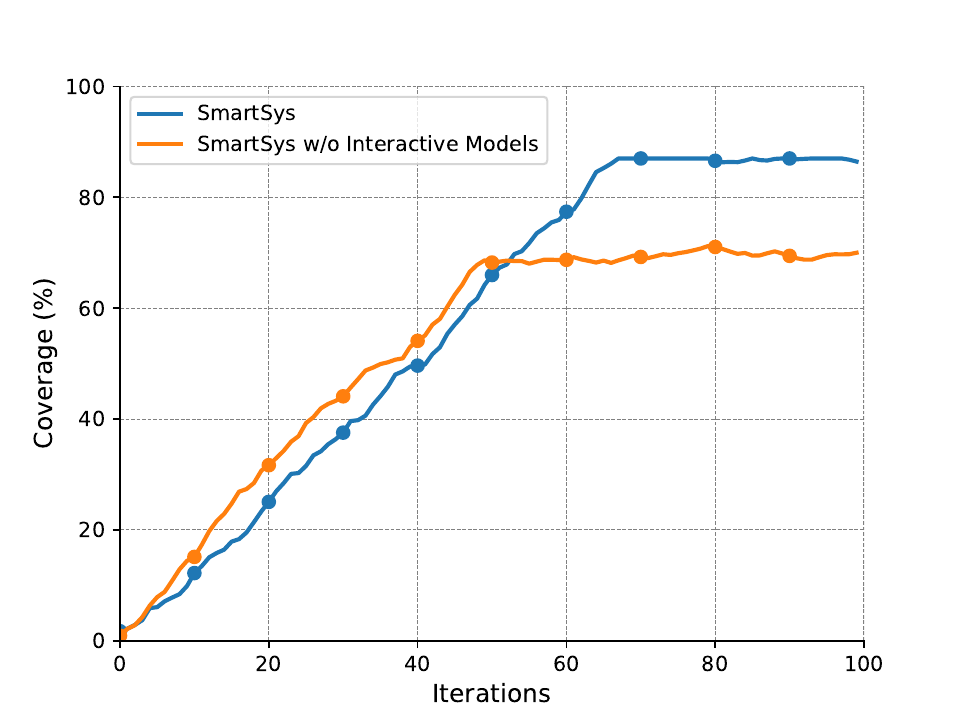} % Replace with your image file
        \label{fig:model_first_bench}
    }
    \hfill
    \subfloat[loop and non-linear constraints benchmark result]{%
        \includegraphics[width=0.3\textwidth]{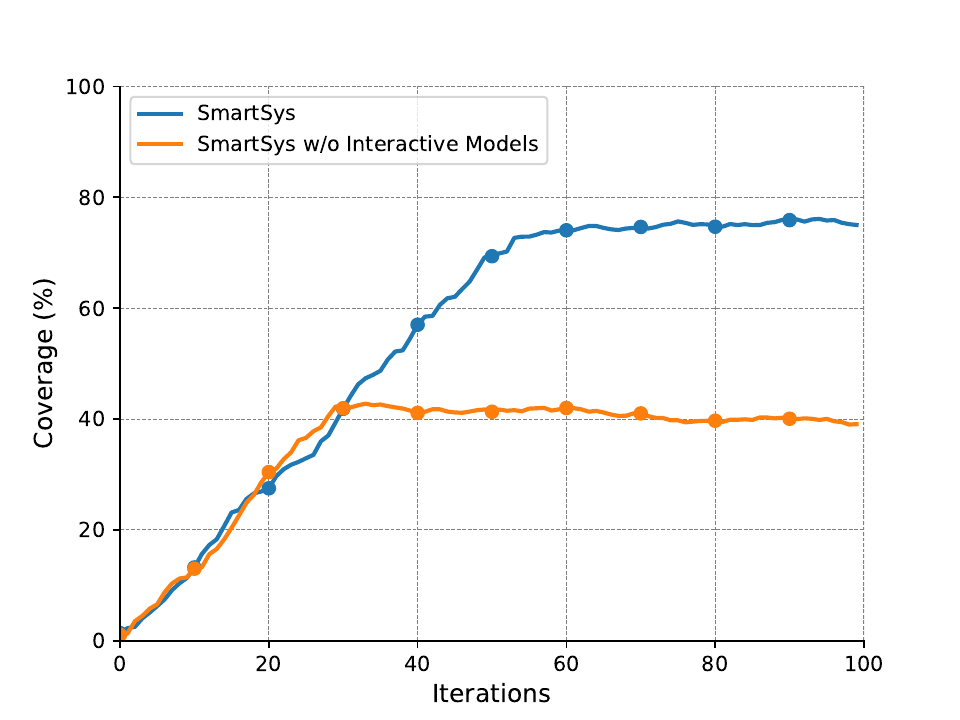} % Replace with your image file
        \label{fig:model_second_bench}
    }
    \hfill
    \subfloat[specific function invocation order benchmark result]{%
        \includegraphics[width=0.3\textwidth]{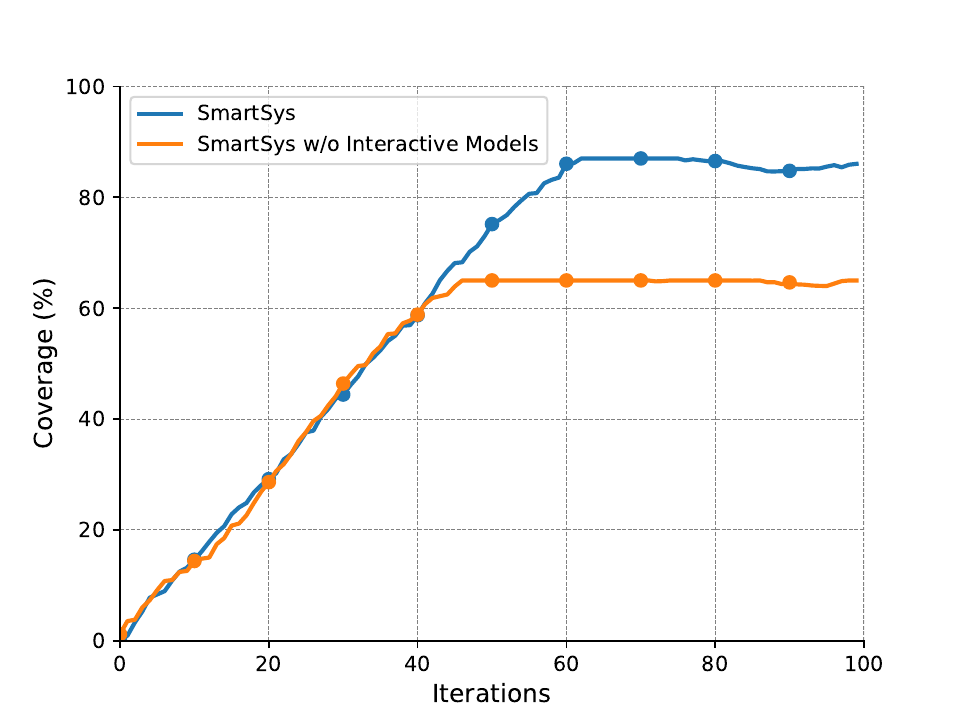} % Replace with your image file
        \label{fig:model_third_bench}
    }
    \caption{The effect of the hallucination suppression algorithm and interactive forecast and generator models on the three benchmarks of D1: The top row shows the effect of the hallucination suppression algorithm; the bottom row shows the effect of interactive models.}
\end{figure*}

\subsection{RQ2: Effect of Individual Components} 
    \label{full_fig:components_cov}
\label{subsec:RQ2}

\vspace{.1cm}\noindent\textbf{The Effect of Hallucination Suppression Algorithm.} Figs. ~\ref{fig:hallucination_first_bench}-\ref{fig:hallucination_third_bench} demonstrates the effect of the hallucination suppression algorithm, with noticeable coverage improvement in Fig.~\ref{fig:hallucination_third_bench}.  The algorithm alone obtained between 4\% and 30\% code coverage improvement. The hallucination suppression algorithm has the most salient effect on the third benchmark, because specific function invocation ordering requires domain-specific transactional knowledge that is inherently challenging for code-based search heuristics.
This result shows that hallucination suppression algorithm is a significant contributing factor to \sys' success.

\vspace{.1cm}\noindent\textbf{The Effect of Interactive Models.} Figs.~\ref{fig:model_first_bench}-\ref{fig:model_third_bench} demonstrates the effect of the interactions between the forecast and generator models. On the three benchmarks, interactive models improved coverage by 11\%, 32\%, 19\% respectively. In addition to coverage improvement, we observed that without interactive models, \sys saturated at lower coverage point during early iterations. By comparison, with interactive models, \sys demonstrated sustained 
coverage improvement over long ($\geqslant$50) iterations. This result shows that interactive models aid both scalability and coverage of \sys. Additional comparisons of the hallucination suppression algorithm and interactive models with other tools are in Appendix~\ref{subsec: additional_eval}. 

\vspace{.1cm}\noindent\textbf{Analysis on Models' Internal Interactions.} To understand how the forecast and generator models interact with each other, we sampled ten contracts from D1 and reported the breakdown in Tab.~\ref{table:cov_internal}. We examined the execution traces of \sys and reported a granular contribution to coverage improvement by \sys' forecast and generator models.

Tab.~\ref{table:cov_internal} shows that on average of five runs of the sampled contracts, \sys collected 23 uncovered functions from the benchmark contracts, totaling 57 lines of code uncovered after the fuzzing campaigns. The forecast model predicted 15 (65\%) of the 23 uncovered functions to be handled by the generator model. The remaining functions were left for the concolic execution engine. This decision led to 
coverage improvement by 49 lines. Equivalently, the generator model reduced uncovered lines by 86\% .
%\junfeng{should also show some internal metrics, such as how often is the forecast model predicts symbolic execution, and how successful is that. also, how many tests are generated, etc.}
\begin{table}
\caption{Analysis on models' interactions internally. \#Func.: the number of uncovered functions extracted from the fuzz scanner. \#Pred.: the number of forecasts on using the generator model for extracted uncovered functions. Lines Uncov.: the total number of lines not covered from extracted uncovered functions. Cov. Impr.: coverage improvement from running the generator model. LoC: lines of code in a contract.}
\label{table:cov_internal}
\footnotesize
\setlength{\tabcolsep}{2.5pt}
%\centering
\renewcommand{\arraystretch}{1.2}
\begin{tabular}{l|l|l|l|l|l}
\hline
Contracts          & \#Func. & \#Pred. & Lines Uncov. & Cov. Impr. & LoC   \\ \hline
\rowcolor[gray]{.9}
Flags              & 2       & 2       & 5            & 5          & 56    \\
CryptoBets         & 4       & 1       & 3            & 3          & 848   \\
\rowcolor[gray]{.9}
Migrations         & 0       & N/A     & 4            & N/A        & 23    \\
Crowdsale\_Complex  & 2       & 2       & 6            & 6          & 82    \\
\rowcolor[gray]{.9}
DoubleSha          & 1       & 0       & 4            & 2          & 65    \\
SpankChain         & 4       & 2       & 12           & 12         & 1319  \\
\rowcolor[gray]{.9}
MCR20              & 3       & 1       & 6            & 4          & 597   \\
MiniMeTokenFactory & 2       & 2       & 6            & 6          & 605   \\
\rowcolor[gray]{.9}
Lottery            & 2       & 2       & 5            & 5          & 208   \\
BasicToken         & 3       & 3       & 6            & 6          & 285   \\ \hline
\rowcolor[gray]{.9}
\textbf{Total}     & 23      & 15      & 57           & 49         & 4,088 \\ \hline
\end{tabular}
\end{table}

\subsection{RQ3: Bug Detection} 
\label{subsec:RQ3}
% \begin{scriptsize}
 \mycode
\begin{lstlisting}[float,floatplacement=H,basicstyle=\fontsize{8}{9}\selectfont, caption={Common bug patterns missed by symbolic execution with manually crafted heuristics and hybrid fuzzing.}, label={bug_type},numbers=left, xleftmargin=2em]
function update(uint price) internal {
    if (price*price*price<=1096) {
        return price;
    }
    return price + 10;   
}

function check(uint price) external {
    if (update(price)==100){
        //bug;
    }
    ...
}
\end{lstlisting}
% \end{scriptsize}
\begin{figure}[!t]
\centering
\includegraphics[width=\linewidth]{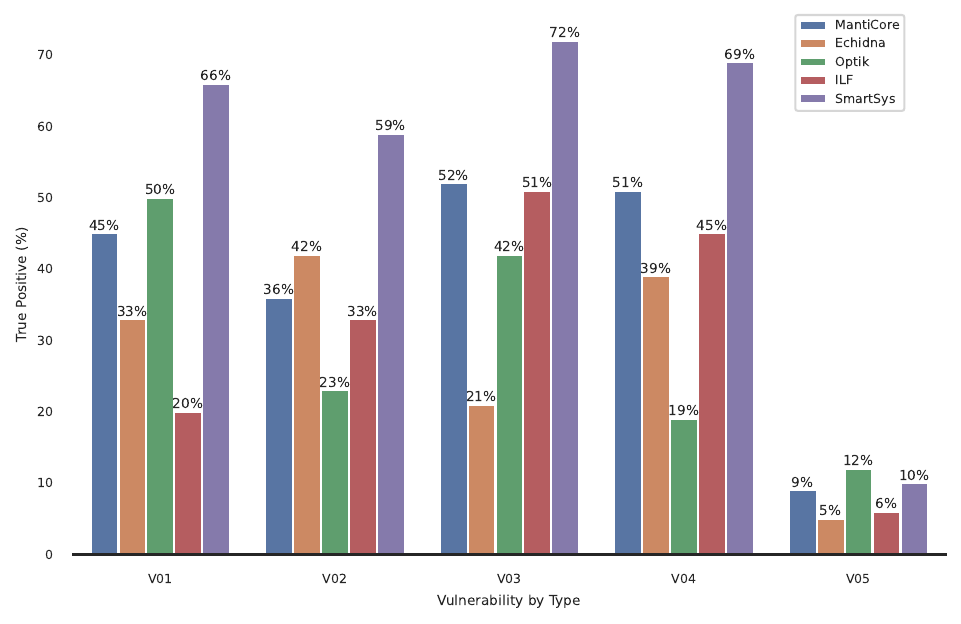}
\caption{Bug detection results on Immunefi's \cite{Immunefi} top five zero-day vulnerabilities. V01: improper input validation. V02: Incorrect Calculation. V03: Oracle/Price Manipulation. V04: Weak Access Control. V05: Signature Malleability.}
\label{fig:bug}
\end{figure}

\vspace{.1cm}\noindent\textbf{Ground Truths.} Ground truths are based on labeled bugs and buggy locations in dataset No.3 (D3) from Immunefi \cite{Immunefi} and Code4Rena \cite{code4rena} auditing reports. The audited reports accounted for 43 out of 60 in our ground truth labels. The remaining ground truths were obtained from SmartBug Curated Dataset \cite{smartbugs} and the benchmarks of Echidna and EFCF, which provided annotated vulnerabilities. 

During evaluation, if a tool reported bugs that did not conform to annotated ground truths, we manually inspected the bugs first. Then we created a forked mainnet environment using Foundry \cite{foundry} to reproduce the reported bugs. If our reproducing efforts confirmed the existence of reported bugs, we treated the reported bugs as ground truth. After further inspection, we confirmed 72 bugs from 60 contracts in total, as some contracts contained multiple bugs.

The tools in Fig.~\ref{fig:full_coverage} were selected for bug detection evaluation, because these tools were originally designed for detecting a broad range of bugs. It is worth noting that the tools did not always report the specific bug type, so we had to make some assumptions to ensure fair evaluation across different tools: when a tool generated test cases or exploits that covered the buggy code in a vulnerable contract, we treated such an instance as a successful detection. We decided not to select \smartestnq\xspace and \ethbmcnq\xspace after devoting four weeks to studying and debugging these symbolic reasoning tools based on the following learned lesson: while they were wildly successful in detecting specific classes of bugs such as integer overflow/underflow and gas leakage as intended, they were not designed for detecting common ``deep bugs'' or zero-day vulnerabilities from recent auditing reports. 

\vspace{.1cm}\noindent\textbf{True Positive Analysis.} Fig.~\ref{fig:bug} summarizes the correctly identified bug results by bug types. 
The five bug types in Fig.~\ref{fig:bug} are the most popular bugs frequently exploited~\cite{popularbugs}. The top performing areas of \sys were detecting V01: improper input validation, V02: incorrect calculation, V03: price manipulation, and V04: weak access control. \sys detected more than 80\% of true positive bugs in these areas. This performance was driven by \sys's generator model that generated effective input values and invocation orders for V01, V03, and V04. Additionally, \sys's optimized concolic execution showed effectiveness for solving constraints for V02. On the most challenging V05: signature malleability, \sys still outperformed similar tools and found 10\% true positives. 

By comparison, existing tools detected fewer true positive bugs than full \sys by up to 22\%. The experiments revealed three key performance bottlenecks that \manticorenq, \optiknq, \echidnanq, and \ilfnq\xspace from achieving better results. First, existing tools did not reason about transactional logic embedded in smart contract code, thereby missing a large portion of V02 and V03 bugs that were rooted in specific function invocation order. Second, existing symbolic execution tools similar to \manticorenq\xspace and \ilfnq\xspace prematurely aborted promising paths (paths that can lead to crashes and bugs), because those paths did not fit into pre-defined heuristics. Third, as \ilfnq's\xspace authors graciously acknowledged in their paper, tools such as \ilfnq's\xspace and \echidnanq\xspace could not reason beyond two inter-contract communications. 

Listing~\ref{bug_type} illustrates the code pattern where existing tools failed to explore the paths leading to the bug within the given time budget. Baseline concolic execution tools \cite{teether, ethbmc} did not cover lines 1-7 and aborted the paths leading to the buggy line 11, because an internal function such as \texttt{update(...)} did not fit into their search heuristics of critical paths. These critical paths were broadly defined as external functions that contain \texttt{CALL}, \texttt{DELEGATECALL}, or \texttt{SELFDESTRUCT} instructions. Such manually crafted search heuristics missed potential constraints and paths that led to critical vulnerable sections of smart contracts. Although \ilfnq\xspace learned to fuzz from symbolic executed paths, it still missed the buggy line, because the non-linear constraints at line 2 caused solver timeout.

\vspace{.1cm}\noindent\textbf{False Positive and Negative Analysis.} \sys, \manticorenq, and \ilfnq\xspace use concolic (or symbolic in the case of \ilfnq\xspace) execution engines to filter out false positive test cases by modeling real transaction executions in Ethereum environment, and as a result, none of them suffered from false positives. We also inspected the false negative results (defined as missed bugs) on D3 dataset. \sys had lower false negatives than the others on all bug types and \optiknq\xspace on four bug types. The false negative results were largely driven by missed coverage on critical buggy lines in large contracts with convoluted code patterns. \sys derived a higher number of fuzz targets for specific function invocation ordering and from the forecast and generator models. These test cases reached deeper code states than test cases from the baseline modules alone. That was why \sys had lower false negatives than other tools. 

An experiment observation worth mentioning is that \sys also completed the highest number of analyses on large contracts ($\geqslant$ 500 lines of code), successfully analyzing 95\% of contracts under test. By comparison, existing tools frequently had errors and timeouts on medium and large contracts. As a result, existing tools' successful analysis rate was $\sim$ 90\% or below. Timeouts were particularly common for \ilfnq\xspace and \manticorenq, because it treated each branch and state equally and were stuck in loops. Compiler incompatibility was the most common error of existing tools. 

\subsection{RQ4: Seeds (Test Case) Quality} 
\label{subsec:RQ4}
\begin{figure}[!t]
\centering
\includegraphics[width=0.8\linewidth]{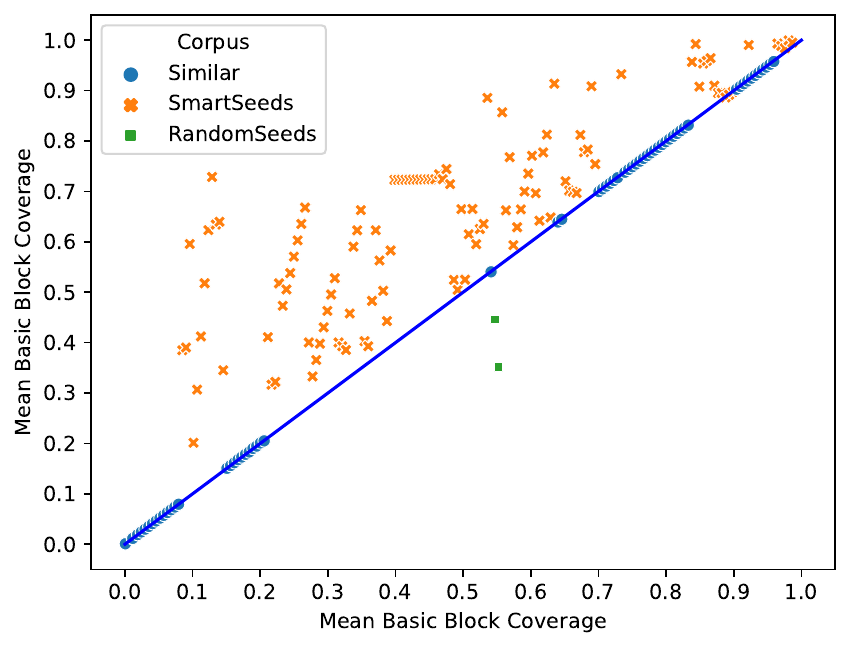}
\caption{Mean basic block coverage reported by EFCF \cite{efcf} from seeds generated by randomly mutated test cases and \sys (SmartSeeds) on D1 evaluation dataset. \colorbox{yellow}{Yellow}: \sys outperformed random seeds. \colorbox{blue}{Blue}: \sys performed similarly random seeds. \colorbox{green}{Green}: Random seeds outperformed \sys.}
\label{fig:seeds_eval}
\end{figure}

We studied the effectiveness of \sys-generated seeds on high throughput fuzzers such as AFL++ adapted  EFCF \cite{efcf}. Using the Solidity to C++ translation infrastructure built by Rodler et al.\cite{efcf}, we ran \sys on 82 randomly sampled projects from dataset D3. Fig.~\ref{fig:seeds_eval} shows that \sys facilitated better coverage by up to 20\% on 52 (63\%) projects (marked by yellow) and similar coverage (marked by blue in Fig.~\ref{fig:seeds_eval}) on 48 (58\%) projects. We investigated the two projects where \sys covered fewer code than random seeds (marked by green). On those two projects, \sys generated incorrect seeds when hallucination algorithm reached specified iterations at 30.

%The key insights into dynamic analysis (concolic execution) and learning-based (foundation models) test case generation are this: unlike concolic execution offers that performance guarantees, foundation models can introduce unsound and inconsistent test cases while improving coverage and detection performance. Fine-tuning and in-context prompting can be useful strategies to mitigate the loss of consistency and decreased precision. 

\section{Related Works}
\label{subsec:related_work_and_technique}
%\input{Tables/related_work}

%\junfeng{do we need this section here?}
While a wide range of techniques are available for smart contracts testing, achieving high scalability and minimizing hallucination automatically remains an open challenge. We analyze existing techniques' advantages and disadvantages. Our key insight is that choosing the right technique prior to heavily investing computational resources can combine the best of all worlds without sacrificing performance.

\vspace{0.1cm}\noindent\textbf{Foundation Models for Code.} Foundation models have shown great promise in assisting code generation and code understanding tasks in recent works \cite{smartinv, llm4fuzz, assemble, codamosa, codellama}. But no prior work has tapped into foundation models' forecasting ability to unite with the power of dynamic analyses and foundation models' code generation ability. This work aims to close the gap in smart contract security space by exploring foundation models' forecast and code generation ability. It demonstrates a prospect of integrating hallucination suppressed foundation models with dynamic analysis.
%\junfeng{ambiguity in this sentence. perhaps swap dynamic analyses with code generation without hallucination.}

\vspace{0.1cm}\noindent\textbf{Fuzz Targets and Fuzzing.} Our fuzz target is a function that accepts input parameters and mutates values for these parameters using the API under test \cite{fuzztarget}. We focus on the pros and cons of smart contracts related tools. Existing fuzzers \cite{fuzzingstateart} usually aim for increasing code coverage by three main optimizations in the smart contract domain: i) GPU or learning based, such as \mau\xspace and \ilf; \maunq\xspace leverages GPUs to parallel and scale up fuzzing.  \ilfnq\xspace learns from symbolic execution to guide fuzzing. ii) Evolutionary algorithm guided, such as \echidna, \sfuzz\xspace and \efcf; iii) hybrid fuzzing that combines fuzzing and symbolic execution, such as \optik. Xia et al. \cite{fuzz4all} correctly acknowledges that solving hallucination remains a worthy challenge in hybrid fuzzing.

\vspace{0.1cm}\noindent\textbf{Concolic (Symbolic) Execution.} Concolic or symbolic execution such as \teether\xspace, \sailfish\xspace and others \cite{multise, solsee} demonstrates superior ability in solving complex constraints that can block fuzzers and in analyzing code beyond those constraints. In an ideal world where memory and time are unlimited, the trade-off is that symbolic execution requires significant time (it can be hours) and memory to achieve full code coverage, due to the well-known path explosion problem. Optimizations trying to mitigate the path explosion problem, such as loop bounding in \teethernq\xspace and paths stitching in \sailfishnq, restrict symbolic execution engines to selected critical paths. As a result, symbolic execution engines' bug detection scope is often severely curtailed by the optimizations for runtime performance.

\section{Conclusion}
\label{sec:conclusion}

\sys is an interactive, self-deciding system that 
% generates hallucination-suppressed fuzz targets to reach deep bugs automatically and scalably.
detects deep bugs in smart contracts automatically and scalably.
\sys uses a fine-tuned model to forecast whether concolic execution or the generator model is the right technique to overcome coverage plateau.
\sys further leverages feedbacks from other tools to generate hallucination-suppressed and bug-revealing fuzz targets.
% \sys integrates the strengths of dynamic analysis and foundation models by forecasting the right technique, concolic execution or fuzzing assisted by the generator model, to overcome coverage plateau. 
Comprehensive evaluation shows that \sys consistently achieves high coverage and detects a wide range of vulnerabilities.

%%% Local Variables:
%%% mode: latex
%%% TeX-master: "main"
%%% End:

\bibliographystyle{plain}
\bibliography{paper}
\section{Appendix}
\label{sec:appendix}

\subsection{Forecast Model Fine-tuning Procedure}
\label{subsec: finetuning_implementation}
\begin{sloppypar}
\vspace{0.1cm}\noindent\textbf{Fine-tuning and Prompting for Forecast.} Our approach is uniquely tailored to fine-tune the model's code understanding ability and thus to improve its forecast accuracy. We used 5-fold cross validation on 121 contracts to construct the training dataset and the holdout test dataset. For each fold on average, we split the contracts into 95 training contracts and 26 holdout test contracts (further details in \S\ref{subsec:three_datasets}). From the training contracts, we extracted 201 training samples from training contracts that did not achieve full coverage from the prior fuzzing stage. 
\end{sloppypar}

The prompts used for fine-tuning are shown below. In the first prompt, we first supply a general context to the model with smart contract's source code. In the second prompt, the model is prompted to learn about the fuzzing results with a pointed direction towards un-covered functions. In the third prompt, the model learns to predict which module (concolic execution or generator model) should be the next test case generator to improve the coverage of  un-covered functions and to find more bugs. \%...\% denotes the content of the five selected features in the training dataset. Each training sample is separated by the embedded special token $<$End of Text$>$. 

\vspace{0.1cm}\noindent\fcolorbox{black}{gray!20}{\begin{minipage}{24em}
Prompt 1: This is the source code: \%smart contract\%.\\

Prompt 2: After fuzzing the above source code, the coverage is \%coverage report\% and detected bugs are \%bug report\%. From the coverage report, the un-covered functions are \%un-covered functions\%. Please understand the un-covered functions. \\

Prompt 3: Let concolic execution be 0 and foundation model be 1. Given the source code and un-covered functions above, the best way to improve coverage and find more bugs is: \%ground truth\%.\\
$<$End of Text$>$
\end{minipage}}\vspace{0.1cm}

\vspace{0.1cm}A learned insight from the prompt design experiments is that the model tends to learn and generalize better with reinforcing statements. Take the second prompt as an example. The reinforcing statement at the end - ``Please understand the un-covered functions'' - keeps the model focused on escaping coverage plateaus of un-covered functions. Without the reinforcing statement, the model rambles on with peripheral information about the smart contract source code and hallucinates with incorrect information. 

\vspace{0.1cm}\noindent\textbf{Inference and Prompting.} After fine-tuning, \sys imports the fine-tuned model for inference. We use similar prompts for inference as for fine-tuning, except that there is no ground truth. By the design of our fine-tuning procedure above, the fine-tuned model is ready to forecast for the next test case module by learning from diverse patterns of un-covered code. 

 During inference, \sys firsts fuzzes the previously unseen contract under test to obtain coverage report and bug report. Then from the fuzzing results, \sys provides contexts to the model with the first two prompts. At the last step of the third prompt, the model is asked to predict next test case generation module. To reinforce the desired format of model's forecast output, \sys adds an additional statement - ``Please generate 0 or 1 only'' - at the end of the third prompt. 

\subsection{Generator Model Prompts}
\label{subsec: genertor_prompts}
The grey box below summarizes the general iterative prompts we use if fuzzing does not achieve full coverage from the start. The first two prompts provide contexts on the function under test and the specific constraints causing coverage plateaus. The third prompt asks for values of each input parameter that can overcome the coverage plateaus. 

\vspace{0.1cm}\noindent\fcolorbox{black}{gray!20}{\begin{minipage}{24em}
Prompt 1: After fuzzing, this is the un-covered function in a smart contract: \%function\%.\\

Prompt 2: The function is un-covered due to the restrictive condition \%constraints\%.\\

Prompt 3: Please fine valid \%input\_parameter\_1, input\_parameter\_2, ..., input\_parameter\_i, ..., input\_parameter\_n\% that can pass the previous given restrictive condition \%constraints\%. Please return the values of each \%input\_parameter\_1, nput\_parameter\_2...input\_parameter\_n\%  only.\\
\end{minipage}}\vspace{0.1cm}

\subsection{Extended Optimized Concolic Execution}
\label{subsec: extended_concolic_execution}
\begin{figure*}[!t]
    \centering 
    \subfloat[Code to generate test cases for]{
         \label{subfig:code}
    \includegraphics[width=.45\linewidth]{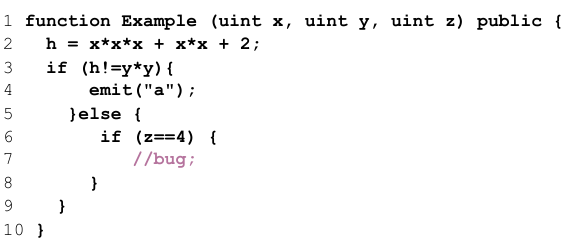}}
    \subfloat[Constraints collecting and solving]{
     \label{subfig:state}
    \includegraphics[width=.45\linewidth]{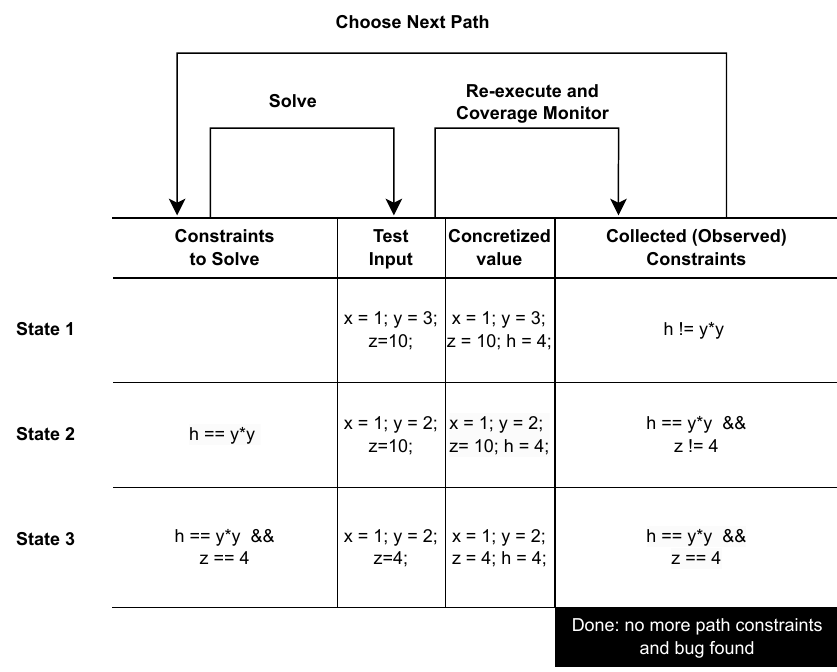}}
        \caption{Concolic execution running example.}
    \label{fig:constraints}
\end{figure*}

\vspace{0.1cm}\noindent\textbf{Ethereum Environment Modeling.} What differentiates our modeling from prior symbolic execution works \cite{ethbmc, teether} is that we model precise, transaction-specific EVM environment, including gas usage and cost \cite{gas}, fallback hooks, inter-contract communications, and inter-transaction communications. The benefits of precise modeling are to reduce \sys's false positive and false negative results and to generate practical test cases. 

We model the environment as an abstract EVM World $\Pi$ that holds transaction objects and reasons about the execution context of a contract. The EVM World $\Pi$  takes a list of externally owned accounts, a contract runner for a list of deployed contract addresses, a runtime stack that includes currently executed function calls with associated contract addresses, a transaction queue that includes a list of transactions to be executed, and a fall back monitor to execute fall back functions when required. When execution is completed, $\Pi$ returns a set of halting instructions, such as \texttt{STOP} and \texttt{RETURN}. 

We model external contract calls, because inter-contract communications are common and such modeling enables the detection of attacks rooted in inter-contract calls \cite{confusum, sailfish}. If a transaction invokes external contracts, $\Pi$ can also return external call instructions, such as \texttt{CREATE} and \texttt{CREATE2}. 

Unlike traditional software, executions of smart contracts' code not only dependent on input, but also the information on blockchain. To capture the blockchain environment, we define an account to be a tuple \emph{a} = (\textsf{address}, \textsf{balance}). The \textsf{address} is an Ethereum address of a given account and the \textsf{balance} is the initial balance of an account in WEI, the smallest unit of Ether. Additionally, 
we define a transaction as a tuple \emph{t} = (\textsf{function\_call}, \textsf{call\_data}, \textsf{delay}, \textsf{gas}, \textsf{gas\_price}, \textsf{source}, \textsf{destination}, \textsf{value}). Given a contract under test, \textsf{function\_call} is the name of the function to be executed. \textsf{call\_data} are symbolic input variables of \textsf{function\_call}. \textsf{delay} is the network delay of each transaction. \textsf{gas} and \textsf{gas\_price} are the amount of gas and gas price for that transaction. \textsf{source} and \textsf{destination} are the addresses of transaction sender and recipient. \textsf{value} is the WEI value attached to the transaction. A contract can have a list of transactions with respect to different function calls and each transaction can be concolicly executed by multiple states.

\vspace{0.1cm}\noindent\textbf{State Representation.} Abstractly, each state can be represented as \emph{s} = (\textsf{path}, \textsf{input}, \textsf{pc}, \textsf{$\Phi$}, \textsf{$\phi$}), \textsf{path} being the explored path leading to the current state; \textsf{input} being the concrete test input being executed; \textsf{pc} being the program counter of the next instruction;  \textsf{$\Phi$} being a list of solved constraints given the current \textsf{path};  \textsf{$\phi$} being the constraint to be solved. Take Fig.~\ref{subfig:code} as an example. Our concolic execution uses symbolic expressions for function \texttt{Example}'s input parameters \texttt{x}, \texttt{y}, \texttt{z} and memory locations at the beginning. The engine executes one path at a time to collect to-be-solved constraints. For example, at the start, \texttt{x} = \texttt{$x_0$}, \texttt{y} = \texttt{$y_0$}, \texttt{z} = \texttt{$z_0$}, where \texttt{$x_0$}, \texttt{$y_0$}, and \texttt{$z_0$} are symbolic values. When code statements at lines 1-4 are concolicly executed in Fig.~\ref{subfig:code}, \textsf{path} is 1, 2, 3, 4. Concretized \textsf{input} can be \texttt{x} = 1, \texttt{y} = 3, \texttt{z} = 10. Conretized values can come from corpus data generated by fuzzing or solved values during symbolic evaluation. \textsf{pc} is 5. \textsf{$\Phi$} is observed constraint \texttt{h != y*y} and \textsf{$\phi$} is to-be-solved constraint \texttt{h == y*y} at line 5.

\vspace{0.1cm}\noindent\textbf{Query Optimization.} We solve constraints using Z3 solver \cite{z3}. To reduce overhead, we want to simplify each query and minimize querying the solver as much as possible. As widely documented, constraints collecting and solving account for the lion's share of performance bottleneck \cite{Klee, ucklee}. Smart contracts are no exception. For our running example Fig~\ref{subfig:code}, without any query optimization, the 10-line code snippets took existing symbolic/concolic execution frameworks \cite{ethbmc, teether, smartian} 10.52 seconds (55.33\%) on average to collect and solve the constraints out of 19.01 seconds end-to-end execution. The overhead is even more salient for large contracts ($\geqslant$ 800 lines). For instance, it took a state-of-the-art\cite{teether} 2,632 seconds, or about 44 minutes, to collect and solve constraints for the 890-line LedgerChannel contract\footnote{https://etherscan.io/address/\\
0xf91546835f756da0c10cfa0cda95b15577b84aa7\#code}. Given the limitations, we spent substantial engineering effort to optimize constraints solving by concretization and snapshots. 

By implementing the DART-style engine for smart contracts, we inherit the benefits of concretizing implied variables for non-linear constraints. Fig.~\ref{subfig:state} shows the constraints collecting and solving process, which is standard by DART \cite{dartconstraint}. The benefits of DART-style concolic execution lie in its ability to efficiently solve non-linear constraints similar to line 2 in Fig.~\ref{subfig:code}. At the else branch at line 5, the to-be-solved constraint is \texttt{h==y*y}. Without concretization, the solver would be required to solve the non-linear constraint \texttt{(x*x*x + x*x + 2)==y*y}, which is known to be a hard and expensive problem \cite{combining, improving}. When \texttt{h} is concretized as 4, this transforms the constraint \texttt{h==y*y} into solving \texttt{y} for \texttt{y*y}==4. This optimization is important for smart contracts, because non-linear constraints are prevalent in asset swapping and auction, where complex mathematical formulas are common.

Another optimization is to take snapshots on solved constraints to avoid repeatedly recreating the EVM world $\Pi$ for each transaction. When analyzing a contract, we set up multiple accounts interacting with the conntract to simulate a chain of transactions $t_1$...$t_i$. For example, Fig.~\ref{subfig:state} represents a single transaction $t_i$ invoking the \texttt{Example} function. We take a snapshot of the EVM world associated with transaction $t_i$. As a result, when $t_i$ is invoked again in different function call invocation order, the concolic execution engine uses the snapshot of $t_i$ instead of creating the entire EVM world again. 

Admittedly, the snapshot solution makes a trade-off between improved runtime performance (eliminating re-execution of same inputs and solving same constraints) versus increased memory cost (caching snapshots of EVM World). This is a worthwhile trade-off in analyzing smart contract transactions, because snapshots reduce runtime overhead by up to 1,091 seconds (49.23\%) and improve coverage by up to 9.5\% on large contracts. In the meantime, we only increased the memory limit from 5GB to 10GB to accommodate snapshots caching, with 25kB per snapshot on average. 

\vspace{0.1cm}\noindent\textbf{Precisely Handling Keccak Hashing.} Keccak hashing is very common in smart contracts, as the EVM includes a dedicated Opcode \texttt{SHA3}. Given a byte size \texttt{S} and memory offset, \texttt{KECCAK256} hash computes the hash of \texttt{S} bytes at the beginning of the target offset. \texttt{KECCAK256} hash is known to be a long-standing challenge for symbolic execution \cite{sha3}. Similar to \cite{confusum, ethbmc}, we concretize the byte size \texttt{S} and the buffer at the target offset when a \texttt{KECCAK256} hash constraint is observed during concolic execution. Given concretized \texttt{S} and target offset,  we calculate the value of the \texttt{KECCAK256} hash. The we assign the calculated value to the variable in the \texttt{KECCAK256} hash operatation. 

\vspace{0.1cm}\noindent\textbf{Coverage Guided Search.} Our search heuristic is to schedule states that will likely cover new code. We have built real-time coverage monitor that measures three types of coverage  during test input re-execution: paths, instructions, and statements. The coverage monitor keeps track of the un-covered instructions. The heuristic prioritizes the scheduling of states that target un-covered instructions.

\vspace{0.1cm}\noindent\textbf{Re-run Test Cases.} Generated test cases are re-run on compiled smart contract byte code natively by a test case replay mechanism that we have built from scratch. In the case of Fig.~\ref{subfig:state}, we show three test cases generated by our concolic execution engine: \{\texttt{x} = 1, \texttt{y} = 3, \texttt{z} = 10\}; \{\texttt{x} = 1, \texttt{y} = 2, \texttt{z} = 10\}; \{\texttt{x} = 1, \texttt{y} = 2, \texttt{z} = 4\}. Each individual test case points to a concolic execution state. The replay mechanism uses the test cases to create actual transaction objects containing the concrete values used in the test cases. \sys then executes the smart contracts using the concrete transaction objects from test cases to update coverage and discover bugs.

\subsection{Additional Evaluation Results}
\label{subsec: additional_eval}

\begin{figure*}[!t]
    \centering
    \subfloat[coverage on benchmark\_1]{%
        \includegraphics[width=0.3\textwidth]{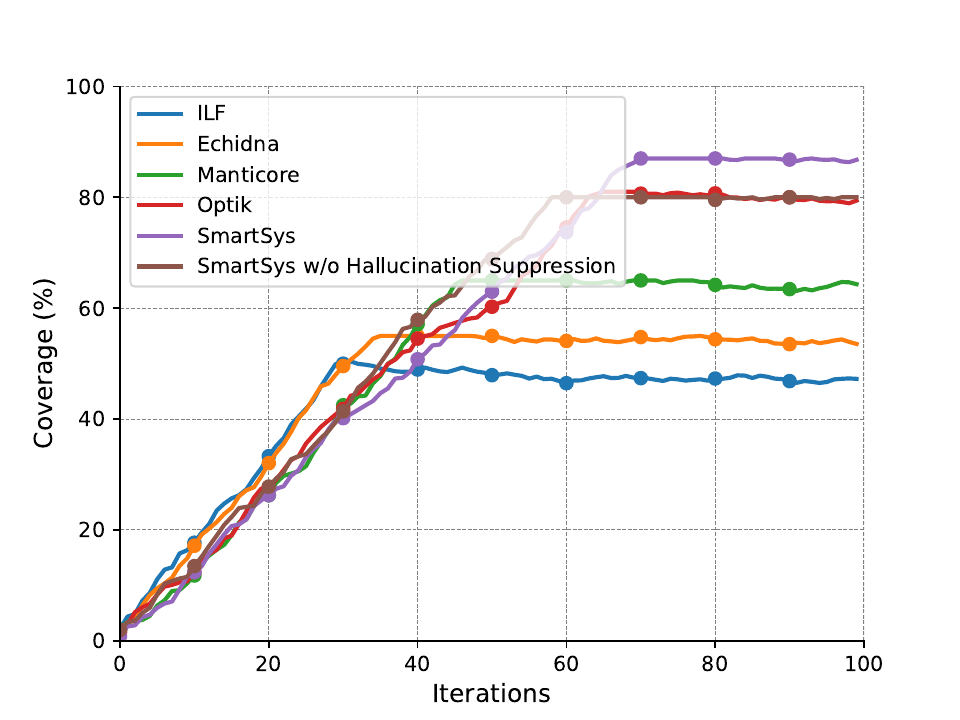} % Replace with your image file
        \label{fig:image1}
    }
    \hfill
    \subfloat[coverage on benchmark\_2]{%
        \includegraphics[width=0.3\textwidth]{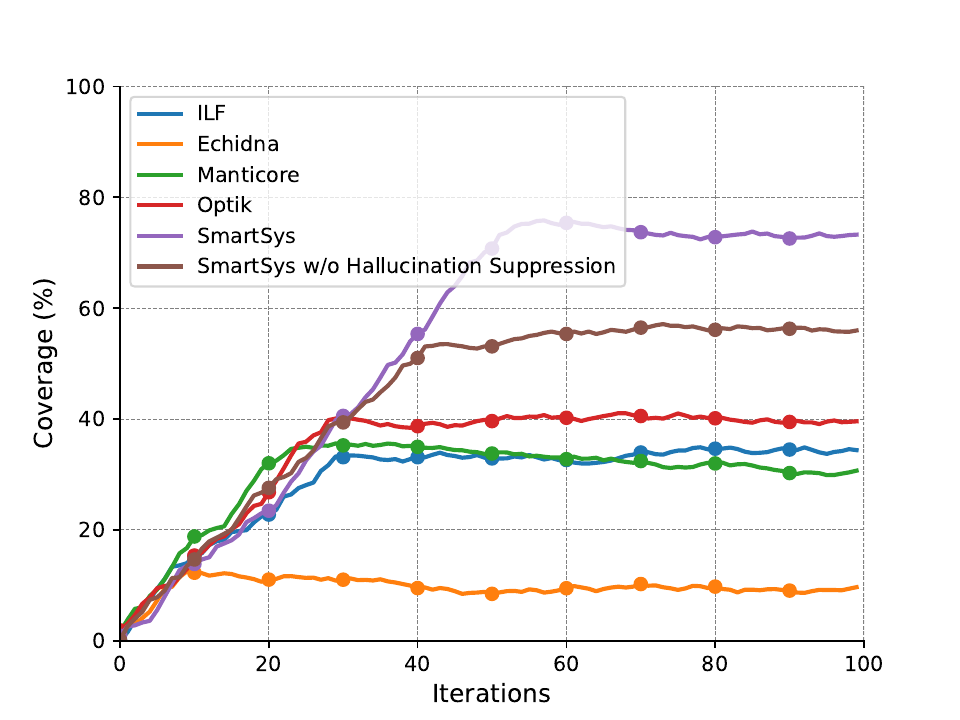} % Replace with your image file
        \label{fig:image2}
    }
    \hfill
    \subfloat[coverage on benchmark\_3]{%
        \includegraphics[width=0.3\textwidth]{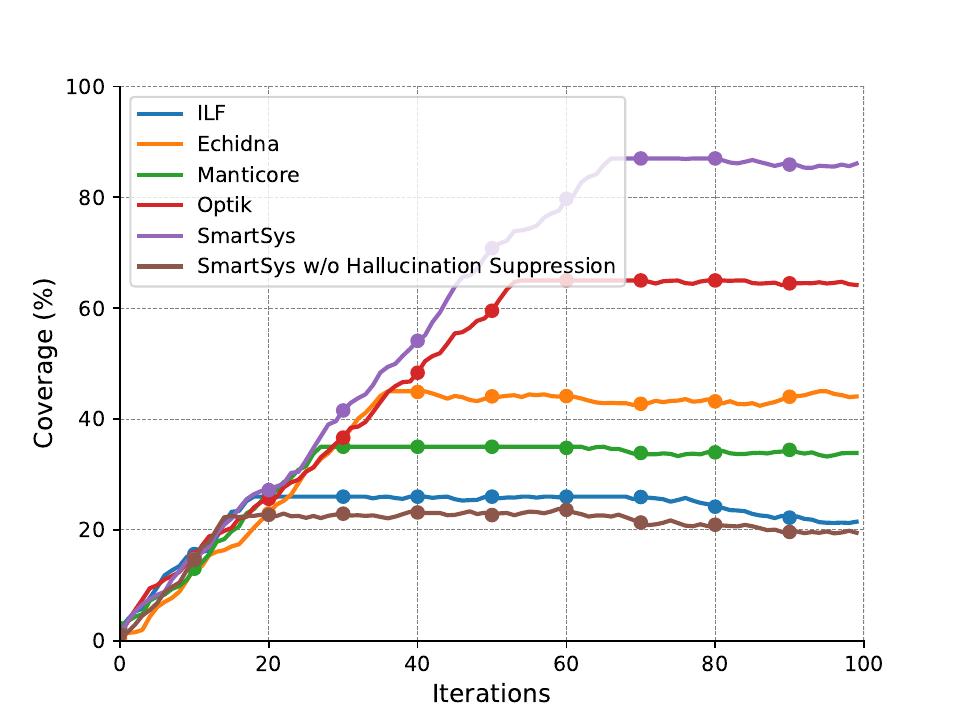} % Replace with your image file
        \label{fig:image3}
    }
    \caption{Overall caption for the three images}
\end{figure*}

\begin{figure*}[!t]
    \centering
    \subfloat[coverage on benchmark\_1]{%
        \includegraphics[width=0.3\textwidth]{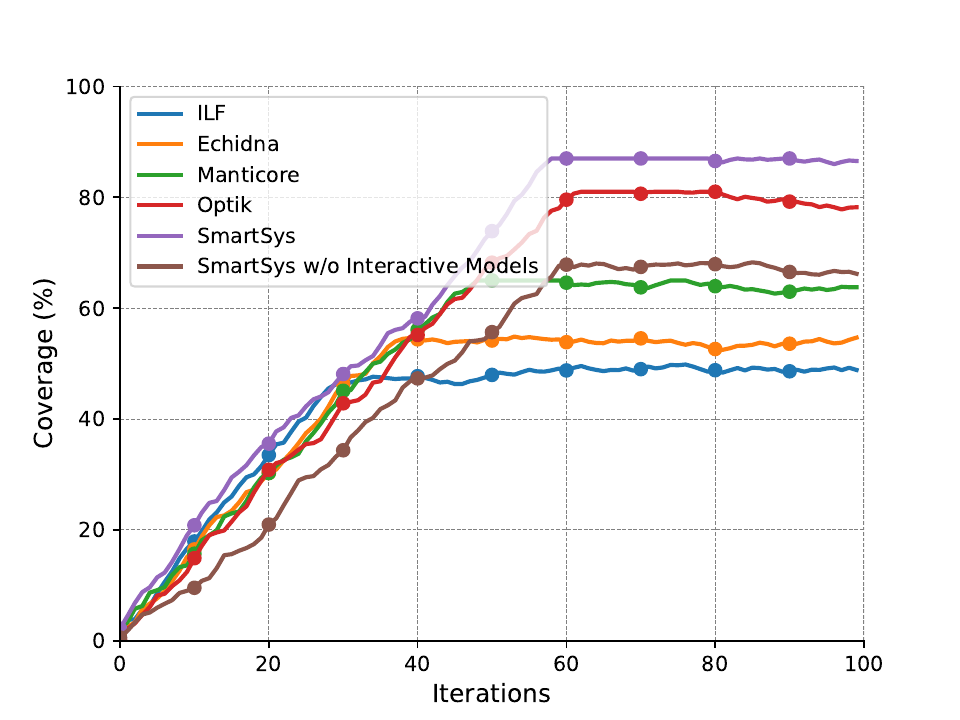} % Replace with your image file
        \label{fig:image1}
    }
    \hfill
    \subfloat[coverage on benchmark\_2]{%
        \includegraphics[width=0.3\textwidth]{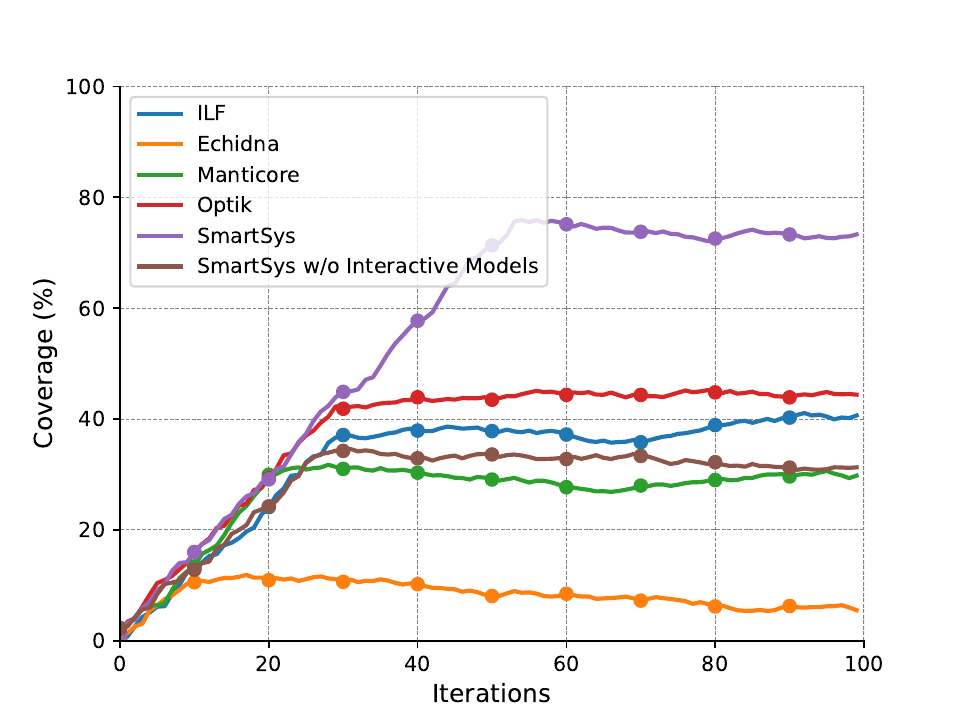} % Replace with your image file
        \label{fig:image2}
    }
    \hfill
    \subfloat[coverage on benchmark\_3]{%
        \includegraphics[width=0.3\textwidth]{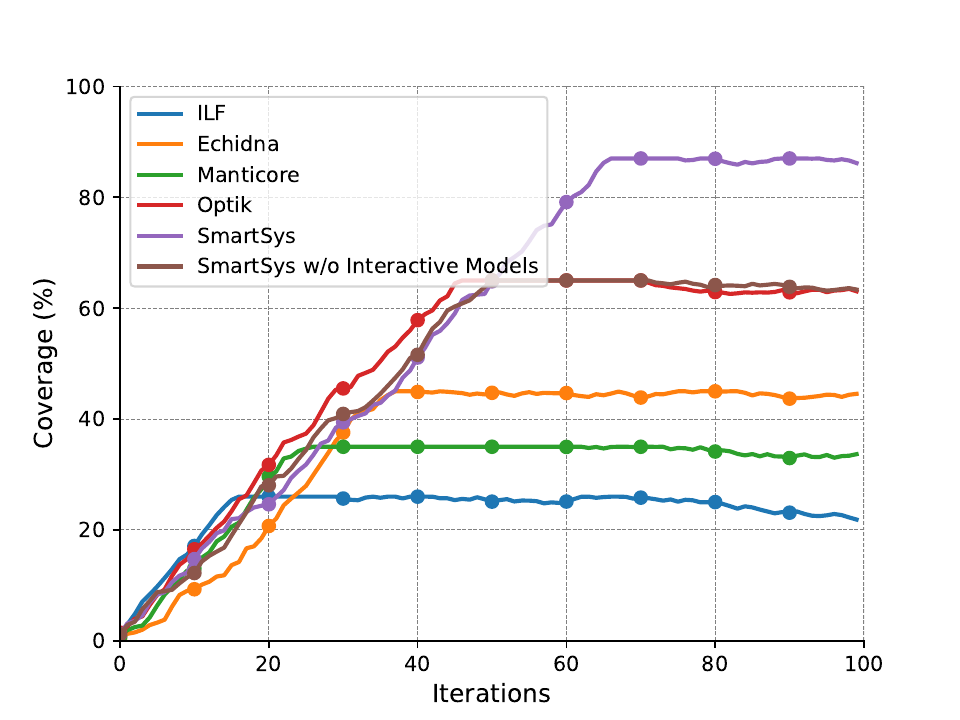} % Replace with your image file
        \label{fig:image3}
    }
    \caption{Code coverage comparisons.}
\end{figure*}

\subsection{RQ4: Ablation Study} 
\label{subsec:ablation}

\vspace{.1cm}\noindent\textbf{Effect of Natural Language.} We studied which models were best suited for forecasting and the effect of multi-modality from fine-tuning on Dataset No.1 (D1) in Tab.~\ref{table:forecast_model}. For fine-tuning, we used 5-fold cross validation, a popular methodology to evaluate models' ability to generalize across unseen data. We split the dataset into 5 randomly sampled folds of similar sizes. We tested fine-tuned  models on each fold. The results demonstrate that CodeGen2 outperformed LLaMA-7B, Falcon-7B, and starCoder by 7\% to 15\% in terms accuracy. However, when we removed natural language by removing comments, documentations, and replacing meaningful variables with ``var'', the accuracy dropped by 8\%. Multi-modality is thus important for the forecast model's performance. 

\vspace{.1cm}\noindent\textbf{Why Fine-tuning or Not.} We observe that without fine-tuning, the forecast model, where \textit{multi-modal understanding} ability is critical, cannot reasonably perform binary predictions on a code snippets. The rationale behind fine-tuning is to equip the model with domain-specific knowledge and train the model to understand what kind of downstream tasks is expected. That observation informed our fine-tuning decision in Tab.~\ref{table:forecast_model}. 

\begin{table}[!t]
\footnotesize
\setlength{\tabcolsep}{8pt}
\centering
\renewcommand{\arraystretch}{1.1}
\caption{\textbf{Fine-tuned candidate models for forecast model. CodeGen2 w/o NL: CodeGen2 without natural language hints.}}
\label{table:forecast_model}

\begin{tabular}{lllll}
\toprule
Model & Acc. & Prec. & Rec.  & F1 \\
 \midrule
% \midrule
\rowcolor{gray!30}
LLaMA-7B \cite{llama} & 0.82  & 0.81 & 0.73 & 0.71\\
% \bottomrule
% \midrule
Falcon-7B \cite{falcon} & 0.81  & 0.81 & 0.70  & 0.75\\
% \bottomrule
% \midrule
\rowcolor{gray!30}
 StarCoder \cite{starcoder} & 0.90 & 0.89 & 0.89  & 0.88\\
\midrule
CodeGen2 \cite{codegen2} w/o NL & 0.89 & 0.90 & 0.90  & 0.89\\
\rowcolor{gray!30}
CodeGen2 \cite{codegen2} & \textbf{0.97} & \textbf{0.96} & \textbf{0.95}  & \textbf{0.96}\\
\bottomrule
\end{tabular}
\end{table}
\begin{table}[!t]
\footnotesize
\setlength{\tabcolsep}{8pt}
\centering
\renewcommand{\arraystretch}{1.1}
\caption{\textbf{Fine-tuned versus vanilla candidate models for generator model on Dataset D2. \#Harness (F): the number of working harnesses generated by a fine-tuned model. \#Harness (V): the number of working harnesses generated by an vanilla model without any fine-tuning.}}
\label{table:generator_model}

\begin{tabular}{lll}
\toprule
Model & \#Harness (F) & \#Harness (V) \\
 \midrule
% \midrule
\rowcolor{gray!30}
LLaMA-7B & 1 & 0\\
% \bottomrule
% \midrule
Falcon-7B & 0  & 1 \\
% \bottomrule
% \midrule
\rowcolor{gray!30}
 StarCoder \cite{starcoder} & 1 & 1\\

CodeGen2 \cite{codegen2} &  0 &  0\\
\midrule
\rowcolor{gray!30}
GPT-4 \cite{codegen2} & \textbf{22}  & \textbf{22} \\
\bottomrule
\end{tabular}
\end{table}

On the other hand, Tab.~\ref{table:generator_model} shows an interesting phenomenon from the experiment: for the generator model, where \textit{code generating} ability is critical, model size played a more important role than fine-tuning. GPT-4, regardless of fine-tuning or not, generated a higher volume of working test harnesses on Dataset D2 than the remaining models in Tab.~\ref{table:generator_model}. This result highlighted that to generate working test harnesses (defined as syntatically correct and compilable test harnesses), the baseline vanilla model selection for the generator model outweighs fine-tuning strategies. 

%\input{Tables/model_paths}

%\subsection{Prompts for Models}
%\label{subsec: prompting}

%%% Local Variables:
%%% mode: latex
%%% TeX-master: "main"
%%% End:

\end{document}